# Path integrals, particular kinds, and strange things

*Karl Friston[1,3], Lancelot Da Costa[1,2]\*, Dalton A.R. Sakthivadivel[3,4,5], Conor Heins[3,6,7], Grigorios A. Pavliotis[2], Maxwell Ramstead[1,3], Thomas Parr[1]*

[1] *Wellcome Centre for Human Neuroimaging, Institute of Neurology, University College London, London WC1N 3AR, UK*
[2] *Department of Mathematics, Imperial College London, London SW7 2AZ, UK*
[3] *VERSES Research Lab, Los Angeles, CA, USA*
[4] *Department of Mathematics, Stony Brook University, Stony Brook, NY, USA*
[5] *Department of Physics and Astronomy, Stony Brook University, Stony Brook, NY, USA*
[6] *Department of Collective Behaviour, Max Planck Institute of Animal Behaviour, Konstanz D-78457, Germany*
[7] *Centre for the Advanced Study of Collective Behaviour, University of Konstanz, Konstanz D-78457, Germany*

*Corresponding author\**

## Abstract

This paper describes a path integral formulation of the free energy principle. The ensuing account expresses the paths or trajectories that a particle takes as it evolves over time. The main results are a method or principle of least action that can be used to emulate the behaviour of particles in open exchange with their external milieu. Particles are defined by a particular partition, in which internal states are individuated from external states by active and sensory blanket states. The variational principle at hand allows one to interpret internal dynamics—of certain kinds of particles—as inferring external states that are hidden behind blanket states. We consider different kinds of particles, and to what extent they can be imbued with an elementary form of inference or sentience. Specifically, we consider the distinction between dissipative and conservative particles, inert and active particles and, finally, ordinary and strange particles. Strange particles can be described as inferring their own actions, endowing them with apparent autonomy or agency. In short—of the kinds of particles afforded by a particular partition—strange kinds may be apt for describing sentient behaviour.

# Introduction

The free energy principle (FEP) describes a simple relationship between the dynamics of a random dynamical system and a description of its behaviour as engaging in inference. The FEP originated in neuroscience as an attempt to describe brain function and behaviour (Friston et al. 2006) and has since been extended to describe several kinds of things in the biological and physical realms (Friston 2013; Friston et al. 2021) through a special kind of mechanics—a Bayesian mechanics—that shares the same foundations with quantum, statistical, and classical mechanics (Friston 2019; Friston et al. 2022). This paper is part of a series of technical papers describing the FEP in progressively simpler and more qualified terms (Friston, 2013; Friston, 2019; Friston et al., 2022).

## A path integral formulation

This paper focuses on a *path integral formulation* of the FEP. The path integral formulation rests upon encoding the dynamics of random dynamical systems by a *Lagrangian* (Graham, 1977b; Seifert, 2012). The Lagrangian plays the role of *self-information* or *surprisal*, scoring the implausibility of a path through state-space. In other words, we will be dealing with probability densities over *paths* or trajectories, as opposed to densities over *states*, which have concerned much of the recent literature on the FEP (Ramstead et al., 2022).

The move from a state-based to a path-based formulation reflects a shift in didactic accounts of the FEP to its operational foundations, expressed in terms of probability densities over paths (Friston, 2010; Friston, 2012; Ramstead et al., 2022). Much of the recent work on the FEP has been concerned with formulating the dynamics of a system in terms of probability densities over states; that is, expressing the probability that certain states or events will occur. Here, we will be concerned with the probability that a system will take a certain trajectory through state space.

The simplifications afforded by the path integral formulation rest on working with densities over paths (as opposed to states). In particular, contrariwise to state-based formulations of the FEP, we make no assumptions about the existence and functional form of a nonequilibrium steady state (Friston et al., 2022; Friston et al., 2021).

## Generalised coordinates of motion

Technically, we will work in *generalised coordinates of motion* (Balaji and Friston, 2011; Da Costa et al., 2021a; Pavliotis, 2014): we augment the traditional state-space of a system with additional degrees of freedom that represent the *n*-order time derivatives of the motion of states. That is, the position, velocity, acceleration, *etc.*, of a particle are represented explicitly as distinct (generalised) states. This simplifies the derivations by replacing a time-series of states (i.e., a path) with a series of time derivatives (i.e., a generalised state). In this setting, the Lagrangian plays the role of an *action*, where paths of least action minimise the Lagrangian of generalised states. Normally, the action and Lagrangian are treated differently, since the Lagrangian is a function of generalised states, and the *action* is a function of paths, usually defined as the path integral of the Lagrangian (Seifert, 2012). These notions coincide in generalised coordinates of motion, as paths are equivalent to a point in generalised coordinates of motion (i.e., a generalised state).

## Particles and particular partitions

The FEP is concerned with self-organisation, which calls for an individuation of 'self' from nonself. We consider a particular partition of (generalised) states necessary to separate the internal states of something (i.e., a particle) from its external states. This partition is defined in terms of sparse coupling among external, sensory, active and internal states. The implications of this particular partition are then unpacked in terms of a variational free energy lemma, which says that the most likely autonomous (i.e., active and internal) paths minimise a free energy functional of Bayesian beliefs about external paths. Crucially, particles with conservative dynamics always pursue paths of least action and therefore minimise variational free energy. The ensuing principle or method of least action is considered in the light of a Bayesian mechanics for describing active particles that possess active states. The autonomous paths of these kinds of particles have a Lagrangian, known as *expected free energy*, which can be decomposed into terms corresponding to expected cost and expected information gain. In this setting, cost is the Lagrangian or surprisal of sensory paths, which defines the characteristic trajectories of a particle. This means that conservative particles comply with the dual optimality principles of Bayesian decision theory (Berger, 2011; Wald, 1947) and Bayesian optimal design (Lindley, 1956; Mackay, 1992). An alternative reading of sensory surprisal is in terms of Bayesian model evidence (a.k.a., marginal likelihood) leading to a description of self-organisation as self-evidencing (Hohwy, 2016).

## Particular kinds

The ensuing formulation leads naturally to a typology of kinds of particles, or 'particular kinds'. We first consider the distinction between active and inert particles, which do and do not have active states, respectively. We then consider a distinction between dissipative and conservative particles, which are and are not subject to random fluctuations, respectively. Finally, we turn to a distinction between ordinary and strange particles, whose active states do, and do not, influence internal states, respectively. This distinction means that the Bayesian beliefs held by strange particles cover (the consequences of) their action. In other words, the active states of strange particles are hidden from internal states and become the latent causes of sensory states that are inferred. This leads to a certain kind of (strange) particle that (looks as if it) believes that it is a conservative particle. In consequence, it actively minimises expected free energy via information and preference-seeking behaviour (Barto et al., 2013; Sun et al., 2011). This sentient behaviour or active inference can be expressed as minimising a generalised free energy functional, which could be regarded as a universal objective function in the design or modelling of agents (Parr and Friston, 2019). This sort of enactive inference (Ramstead et al., 2020) can be read as generalising *planning as inference* (Attias, 2003; Botvinick and Toussaint, 2012; Lanillos et al., 2021).

# The free energy principle

The FEP is a simple account of self-organisation that shares the same foundations with quantum, statistical, and classical mechanics, and leads to a special kind of mechanics—a Bayesian mechanics—that we can use to describe the dynamics of certain 'things' as engaging in inference (Fields et al., 2021a; Friston, 2019; Ramstead et al., 2022). The FEP is derived by paying careful attention to the way 'things' are defined and, thereby, the way in which they are both individuated from—and coupled to—everything else. Before laying down a path integral formulation of the FEP, we summarise the steps in narrative form:

- The FEP addresses the following question: if something exists, in the sense of possessing characteristic states or dynamics, what properties must it possess? To answer this question, it is necessary to define a thing or particle.

- A particle is constituted by internal and blanket states. The blanket states constitute the boundary between the states internal and external to the particle. Mathematically, this means that internal paths are conditionally independent of external paths, given blanket paths.

- This conditional independence means that, for every blanket path, there exists a most likely internal path and a (posterior) probability density over external paths.

- The ensuing synchronisation map from the most likely internal path to the conditional density over external paths can be read as *inference*, in the sense that the most likely internal path encodes a Bayesian belief about external paths.

- This Bayesian interpretation can be made explicit by associating the Lagrangian with a variational free energy; namely, a free energy functional of Bayesian beliefs about external states, given blanket states.

- The internal paths of sufficiently large (i.e., conservative) particles are always the most likely paths; namely, paths of least action, where action is the path integral of a Lagrangian (a.k.a., self-information or surprisal) in the usual way.

- The Lagrangian of the active paths (i.e., blanket paths that can be influenced by internal paths) of conservative particles reduces to variational free energy. This means the autonomous (i.e., active and internal) states of such particles will appear to pursue paths of least action that minimise variational free energy.

- In this case, the Lagrangian of autonomous paths can be decomposed into the expected Lagrangian (i.e. implausibility or cost) of sensory paths minus expected information gain; namely, the reduction of uncertainty about external paths.

- This has the interesting interpretation that autonomous paths will appear to minimise expected cost—where cost is read as the Lagrangian of sensory paths that characterises the particle in question—while maximising expected information gain. This is consistent with the principles of Bayesian decision theory—e.g., expected utility theory (Von Neumann and Morgenstern, 1944)—and optimal Bayesian design (Lindley, 1956), respectively.

- The combination of expected information gain and expected cost has the functional form of an expected free energy.

- Finally, we turn to strange particles: conservative particles whose active paths only influence internal paths vicariously, via sensory paths. Strange particles can be read as inferring their own actions—in addition to the external world—endowing them with apparent autonomy or agency.

In summary, starting with a definition of what it is to be a thing—in terms of Markov blankets—one ends up with a Bayesian mechanics of (certain kinds of) things. Internal paths look as if they are inferring external paths—through minimising a variational free energy. Active paths look as if they comply with the principles of optimal Bayesian design (Lindley, 1956) and decision theory (Berger, 2011)—through maximising expected information gain and minimising expected cost, respectively. Among the particular kinds delineated here, strange things may be apt for describing the sentient behaviour of agents.

The resulting FEP can be read as a variational principle of least action, a gauge theory (Friston et al., 2022; Sakthivadivel, 2022c; Sengupta et al., 2016), the dual to Jaynes' maximum entropy or principle of maximum calibre (Sakthivadivel, 2022a; Sakthivadivel, 2022b) or—in its quantum-theoretic formulation—asymptotically equivalent to the Principle of Unitarity (Fields et al., 2021a).

On this view, the FEP is a first principles account or method that can be applied to any 'thing' or 'particle' in a way that dissolves the bright lines between physics, biology and psychology (Chris Fields, personal communication). Such applications endorse many normative accounts of sentient behaviour and self-organisation. These range from cybernetics to synergetics (Ao, 2004; Ashby, 1979; Haken, 1983; Kelso, 2021); from reinforcement learning to artificial curiosity (Barto et al., 2013; Schmidhuber, 1991; Sutton and Barto, 1981; Tsividis et al., 2021); from predictive processing to universal computation (Clark, 2013b; Hohwy, 2016; Hutter, 2006); from model predictive control to empowerment (Hafner et al., 2020; Klyubin et al., 2005), and so on. We now unpack the narrative arguments above, using standard results from statistical physics and information theory.

# The path integral formulation

*"We do not find obvious evidence of life or mind in so-called inert matter; but if the scientific point of view is correct, we shall ultimately find them, at least in rudimentary form, all through the universe."* (Haldane, 1932)

We begin by describing the setup. We assume that the states of the system—which we will later decompose into a particle and its external milieu—evolve as a random dynamical system in a *d*-dimensional Euclidean space that can be described with a Langevin equation,

$$\dot{x}(\tau) = f(x) + \omega(\tau): \quad 0 \leq \tau \leq T. \tag{1}$$

This equation describes the rate of change of states $x(\tau)$, in terms of their flow $f(x)$ (i.e., a vector field), and some random fluctuations $\omega(\tau)$ with smooth (analytic) sample paths. This setup speaks nicely to the fact that, in biology, fluctuations are often smooth up to a certain order (Vasseur and Yodzis 2004) as they are generated by other random dynamical systems (Friston 2019). We assume that the fluctuations are state-independent, and a stationary Gaussian process[1], which can be read as a consequence of the central limit theorem; i.e., fluctuations should be normally distributed at each point in time. We make no assumptions about the solutions of this equation, other than the flow operator does not change over some relevant time interval. This furnishes an adiabatic approximation (Born and Fock, 1928) to systems in which the parameters of the flow change slowly. Accordingly, we will be concerned with the dynamics over a suitably small period of time and consider the implicit separation of timescales elsewhere (e.g., the separation of fast inference about states and the slow learning of parameters).

Equation (1) can be expressed in generalised coordinates of motion, $\vec{x} = (x, x', x'', \ldots)$, where, to first-order:[2]

$$\left.\begin{aligned} \dot{x} &= x' = f(x) + \omega \\ \dot{x}' &= x'' = \nabla f \cdot x' + \omega' \\ \dot{x}'' &= x''' = \nabla f \cdot x'' + \omega'' \\ &\vdots \end{aligned}\right\} \quad \Leftrightarrow \quad \begin{aligned} \dot{\vec{x}} &= \mathbf{f}(\vec{x}) + \vec{\omega} \\ p(\vec{\omega}) &= \mathcal{N}(\vec{\omega}; 0, 2\Gamma) \end{aligned}$$

$$\mathbf{D} = \begin{bmatrix} 0 & 1 & & \\ & 0 & 1 & \\ & & 0 & \ddots \\ & & & \ddots \end{bmatrix}, \quad \nabla \mathbf{f}(\vec{x}) = \mathbf{J} = \begin{bmatrix} \nabla f & & & \\ & \nabla f & & \\ & & \nabla f & \\ & & & \ddots \end{bmatrix}, \quad \Gamma = \begin{bmatrix} \Gamma_0 & & -\Gamma_0'' & \\ & \Gamma_0'' & & -\Gamma_0^{(4)} \\ -\Gamma_0'' & & \Gamma_0^{(4)} & \\ & -\Gamma_0^{(4)} & & \ddots \end{bmatrix}$$

$$\tag{2}$$

$\nabla f$ is the Jacobian matrix of the flow, **D** is a (derivative) operator that sends each generalised state to the next, **f** is the flow of the system in generalised coordinates of motion, and **J** is the Jacobian matrix of the flow in generalised coordinates. In generalised coordinates of motion, the state, velocity, acceleration, *etc* are treated as separate (generalised) states, whose flow is supplemented with state-independent smooth fluctuations (Da Costa et al., 2021b; Friston, 2008; Kerr and Graham, 2000; Pavliotis, 2014). In summary, we go beyond white noise assumptions about fluctuations[3] and express analytic or smooth fluctuations $\omega(\tau)$ with autocovariance $\Gamma_\tau = \Gamma(\tau) = \mathbb{E}[\omega(\tau)\omega(0)]/2$.[4]

---

[1] In statistical physics, this type of noise is generally referred to as 'coloured noise', the prototypical example of which is the smoothing of white noise fluctuations with a Gaussian kernel.
[2] Going from (1) to (2) corresponds to solving the *stochastic realisation problem* (Mitter et al. 1981; Da Costa et al. 2021). In this instance, (2) is obtained by recursively differentiating (1) and ignoring the contribution of the derivatives of the flow of order higher than one; see (Balaji and Friston 2011) for details. In other words, the expansion is exact when the flow is linear, and it is accurate on a short timescale when the flow is non-linear.

Generalised coordinates of motion can be read as the coefficients of a Taylor series of the trajectory or path, from any point in time, $\tau$:

$$x(t) = x(\tau) + x'(\tau)(t-\tau) + \frac{x''(\tau)}{2!}(t-\tau)^2 + \ldots \quad (3)$$

This means that generalised states correspond to paths. The path integral formulation (Graham, 1977b; Seifert, 2012) considers the probability of a path in terms of its *action* (i.e., negative log probability), which we can associate with the *Lagrangian* (i.e., negative log probability) of generalised states.[5] From the expansion in generalised coordinates in (2), we have that the surprisal of the generalised states is as follows:[6]

$$\mathcal{L}(\vec{x}) \triangleq -\ln p(\vec{x}) = \tfrac{1}{2}[\ln|\boldsymbol{\Gamma}| + \vec{\omega} \cdot \tfrac{1}{2\boldsymbol{\Gamma}} \vec{\omega}] \quad (4)$$
$$\vec{\omega} = \dot{\vec{x}} - \mathbf{f}(\vec{x})$$

This Lagrangian scores the degree to which fluctuations cause the path to *deviate from the path of least action*. The path of least action—i.e., the most likely path—is simply the flow in the absence of deviations, when the random fluctuations take their most likely value of zero. Paths of least action can be interpreted as the preferred trajectories that characterise a system. We will denote these paths—and their generalised states—by boldface, where, from (4):

$$\dot{\vec{\mathbf{x}}} = \mathbf{D}\vec{\mathbf{x}} = \mathbf{f}(\vec{\mathbf{x}}) \Leftrightarrow \vec{\boldsymbol{\omega}} = 0 \Leftrightarrow \vec{\mathbf{x}} = \arg\min_{\vec{x}} \mathcal{L}(\vec{x}) \Leftrightarrow \nabla_{\vec{x}} \mathcal{L}(\vec{\mathbf{x}}) = 0 \quad (5)$$

---

[3] White noise fluctuations have trivial temporal autocovariance structure, i.e., $\mathbb{E}[\omega(\tau + \epsilon)\omega(\tau)] = 0$ if $\epsilon \neq 0$.
[4] The covariance of the fluctuations on the *n*-th order motion then becomes the *n*-th order temporal derivative of the autocovariance function describing their temporal correlation; please see Appendix A.5.3 in (Parr et al. 2022) for details.
[5] In the case of white noise fluctuations, the *action* is the integral of the Lagrangian over paths, where the usual (Stratonovich) form reads (Seifert 2012):

$$\mathcal{L}(x) = \tfrac{1}{2}[\ln|\Gamma_0| + \omega \cdot \tfrac{1}{2\Gamma}\omega + \nabla \cdot f] : \quad \omega = x' - f(x)$$
$$\mathcal{A}(\tilde{x}) = -\ln p(\tilde{x}) = \int_0^T dt\, \mathcal{L}(x) : \quad \tilde{x} = [x(t)] : 0 \leq t \leq T$$

In generalised coordinates of motion, action can be defined as the Lagrangian $\mathcal{A}(\tilde{x}) = \mathcal{L}(\vec{x})$, via the one-to-one correspondence between paths and generalised states through (3).

[6] Where does the term $\tfrac{1}{2}\ln|\Gamma|$ come from? This term is simply the constant of proportionality that makes the Lagrangian a log probability. This term is suppressed in (Seifert 2012), where it is assumed to be constant. The Lagrangian is quadratic in the fluctuations, which follows from the fact that the probability density over fluctuations is Gaussian—and the normalising constant of a Gaussian density involves the (log) determinant of its covariance.

That is, the flow operator **D** produces the most likely flow when the amplitude of random fluctuations goes to zero, which is precisely the path $\vec{\mathbf{x}}$ that minimises the Lagrangian.[7]

Generalised states afford a convenient way of expressing the path of least action as the solution to the following equations (Friston et al., 2010b), which reduce to (5) when evaluated at the path of least action:

$$\nabla_{\vec{x}} \mathcal{L}(\vec{x}) + (\dot{\vec{x}} - \mathbf{D}\vec{x}) = 0 \Leftrightarrow \dot{\vec{x}} - \mathbf{D}\vec{x} = -\nabla_{\vec{x}} \mathcal{L}(\vec{x}) \Leftrightarrow \dot{\vec{x}}(\tau) = \mathbf{D}\vec{x} - \nabla_{\vec{x}} \mathcal{L}(\vec{x}). \tag{6}$$

The first equality resembles a Lagrange equation of the first kind that ensures the generalised motion of states is the state of generalised motion. Alternatively, it can be read as a gradient descent on the Lagrangian, in a moving frame of reference. When the Lagrangian is convex, any solution to this generalised gradient descent on the Lagrangian converges to the path of least action. Equation (6) will be useful later, when recovering paths of least action, in the spirit of generalised Bayesian filters (Friston et al., 2010b).

The uncertainty over paths derives from random fluctuations, which can be expressed in terms of the expected Lagrangian (i.e., differential entropy): from (2) and (4):[8]

$$\begin{aligned}\mathbb{E}[\mathcal{L}(\vec{x})] &= \tfrac{1}{2} \ln |\mathbf{\Gamma}| + \tfrac{1}{2} \mathbb{E}[\vec{\omega} \cdot \tfrac{1}{2\mathbf{\Gamma}} \vec{\omega}] \\ &= \tfrac{1}{2} \ln |e\mathbf{\Gamma}| \triangleq \mathcal{S}(\mathbf{\Gamma}) \triangleq \mathcal{S}\end{aligned} \tag{7}$$

When the amplitude of random fluctuations tends to zero, there is no uncertainty, and the path is always the path of least action. This corresponds to *conservative*—but potentially chaotic—dynamics. The differential entropy above be expressed in terms of continuous entropy, using the limiting density of $N$ discrete points [eq. 12.7 in (Jaynes 2003)] in relation to a (constant) invariant measure $m(\vec{x}) \equiv m : \forall \vec{x}$:

$$\begin{aligned}\mathbb{H}[p(\vec{x})] &= \ln N - D_{KL}[p(\vec{x}) \| m(\vec{x})] \\ &= \mathbb{E}[\mathcal{L}(\vec{x})] + \ln Nm = \mathcal{S} - \mathcal{S}_0\end{aligned}$$

$$\begin{aligned}\lim_{\mathbf{\Gamma} \to 0} \mathbb{E}[\mathcal{L}(\vec{x})] &= -\ln Nm = \mathcal{S}(0) \triangleq \mathcal{S}_0 \\ &\Leftrightarrow \mathbb{H}[p(\vec{x})] = 0 \\ &\Leftrightarrow p(\vec{x}) = \delta(\vec{\mathbf{x}} - \vec{x}) \Leftrightarrow \vec{\omega} = 0\end{aligned} \tag{8}$$

These expressions say that paths of least action (and their associated generalised states) are confined to sets of vanishing Lebesgue measure, where the expected Lagrangian approaches a lower bound: $\mathcal{S}_0 \leq \mathbb{E}[\mathcal{L}(\vec{x})]$. This inequality follows since the limiting density of discrete points is always non-negative, as it is a limit of a discrete entropy—the fact that it vanishes for paths of least action means that the corresponding generalised states occupy a set of measure zero under the invariant measure [Section 12.3 in (Jaynes 2003)]. We will apply these results later to subsets of sparsely coupled states. For example, given a subset $\pi \subset x$ with parents $pa[\pi] \subset x$, we have:

$$
\begin{aligned}
\mathbb{E}_{p(\vec{\pi}|pa[\vec{\pi}])}\left[\mathcal{L}(\vec{\pi} \mid pa[\vec{\pi}])\right] &= \tfrac{1}{2}\ln|\mathbf{\Gamma}_\pi| + \tfrac{1}{2}\mathbb{E}[\vec{\omega}_\pi \cdot \tfrac{1}{2\Gamma_\pi}\vec{\omega}_\pi] \\
&= \tfrac{1}{2}\ln|e\mathbf{\Gamma}_\pi| \triangleq \mathcal{S}_\pi \geq \mathcal{S}_0 \Rightarrow \\
\mathbb{E}\left[\mathcal{L}(\vec{\pi} \mid pa[\vec{\pi}])\right] &= \mathbb{E}_{p(pa[\vec{\alpha}])}[\mathcal{S}_\pi] \geq \mathcal{S}_0 \\
\vec{\omega}_\pi &= \mathbf{D}\pi - \mathbf{f}_\pi(\vec{\pi}, pa[\vec{\pi}]) \\
\mathcal{L}(\vec{\pi} \mid pa[\vec{\pi}]) &\triangleq -\ln p(\vec{\pi} \mid pa[\vec{\pi}]) \\
\lim_{\Gamma_\pi \to 0}\mathbb{E}\left[\mathcal{L}(\vec{\pi} \mid pa[\vec{\pi}])\right] &= \mathcal{S}_0 \Leftrightarrow \mathbb{H}[p(\vec{\pi} \mid pa[\vec{\pi}])] = 0
\end{aligned}
\qquad (9)
$$

---

[7] The gradient of the Lagrangian vanishes at the global minimum given by the path of least action.

[8] Noting that the expected value of the square of fluctuations (i.e., the covariance) is $2\mathbf{\Gamma}$, the term $\tfrac{1}{2\mathbf{\Gamma}} \mathbb{E}[\vec{\omega} \cdot \vec{\omega}]$ (where we have pulled the constant $\tfrac{1}{2\mathbf{\Gamma}}$ out) evaluates to one, or $\ln |e|$. We can then combine logarithms in the usual way to obtain (7).

Here, the second line is obtained as in (7). The last line expresses the fact that, in the absence of random fluctuations on a subset of states $\pi \subset x$, their motion (and associated generalised state) is completely determined by the parents $pa[\pi] \subset x$, at which point the conditional distribution is a Dirac delta with a continuous entropy of zero, and a differential entropy that approaches its lower bound. In other words, in absence of fluctuations on their motion, a subset of states follow paths of least action. This concludes our setup. We now turn to a particular partition of states that defines 'things' and consider how different kinds of things behave.

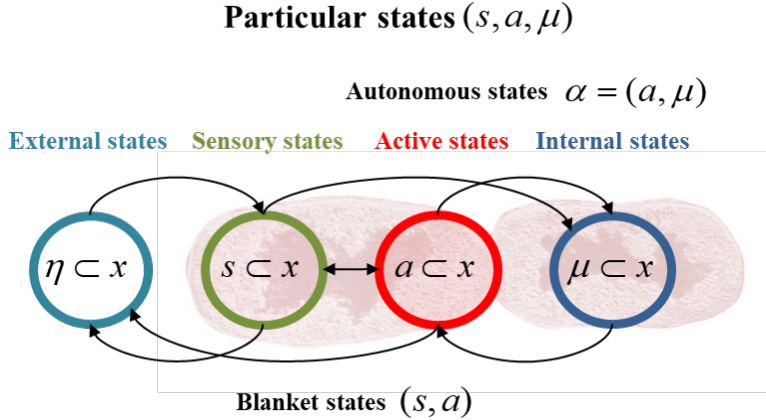

**Figure 1** – *Markov blankets*. This influence diagram illustrates a particular partition of states into internal states (blue) and external states (teal) that are separated by a Markov blanket comprising sensory (green) and active states (red). The diagram shows this partition as it would be applied to a single-cell organism, where internal states are associated with intracellular states, the sensory states become the surface states or cell membrane overlying active states: e.g., the actin filaments of the cytoskeleton. Particular states constitute a particle; namely, autonomous and sensory states—or blanket and internal states. The particular aspect of this coupling is that external states can only influence themselves and sensory states, while internal states can only influence themselves and active states.

## A particular partition and variational free energy

"*How can the events in space and time which take place within the spatial boundary of a living organism be accounted for by physics and chemistry?*" (Schrödinger, 1944).

The free energy principle applies to a particular partition of states $(s,a,\mu) \subset x$ that distinguishes between the states of a particle or 'thing' and its external states $\eta \subset x$, based on their sparse coupling (see Figure 1):[9]

$$\begin{bmatrix} \dot{\eta}(\tau) \\ \dot{s}(\tau) \\ \dot{a}(\tau) \\ \dot{\mu}(\tau) \end{bmatrix} = \begin{bmatrix} f_\eta(\eta,s,a) \\ f_s(\eta,s,a) \\ f_a(s,a,\mu) \\ f_\mu(s,a,\mu) \end{bmatrix} + \begin{bmatrix} \omega_\eta(\tau) \\ \omega_s(\tau) \\ \omega_a(\tau) \\ \omega_\mu(\tau) \end{bmatrix}, \tag{10}$$

where fluctuations $\omega_i, i \in (\eta,s,a,\mu)$ are assumed to be mutually independent. The particular states are partitioned into sensory, active and internal states with particular flow dependencies; namely, external states can only influence themselves and sensory states, while internal states can only influence themselves and active states. From (4), these coupling constraints mean that external and internal paths are independent, when conditioned on blanket paths.

$$\frac{\partial^2 f}{\partial \eta \partial \mu} = 0 \Rightarrow \frac{\partial^2 \mathcal{L}}{\partial \vec{\eta} \partial \vec{\mu}} = 0 \Leftrightarrow \mathcal{L}(\vec{\eta},\vec{\mu} \mid \vec{s},\vec{a}) = \mathcal{L}(\vec{\eta} \mid \vec{s},\vec{a}) + \mathcal{L}(\vec{\mu} \mid \vec{s},\vec{a}) \Leftrightarrow (\vec{\eta} \perp \vec{\mu}) \mid \vec{s},\vec{a}. \tag{11}$$

This is because there are no flows that depend on both internal and external states (and fluctuations are independent).

The conditional independence above distinguishes the dynamics of a particular 'thing' from every 'thing' else. In other words, the statistical separation of internal dynamics from external dynamics rests on the existence of a 'boundary' supplied by a Markov blanket (Pearl, 2009). Why start here? Because the only thing we have at hand is a probabilistic description of the system (in terms of trajectories which correspond to generalised states), and the only way to separate the states of something from its boundary states is in terms of probabilistic independencies—in this instance, conditional independencies.[10] This separation means the internal paths of least action do not depend upon external paths, given blanket paths. From (6) :[11]

$$\begin{aligned}\dot{\vec{\mu}}(\tau) &= \mathbf{D}\vec{\mu} - \nabla_{\vec{\mu}} \mathcal{L}(\vec{\mu},\vec{s},\vec{a},\vec{\eta}) = \mathbf{D}\vec{\mu} - \nabla_{\vec{\mu}} \mathcal{L}(\vec{\mu} \mid \vec{s},\vec{a},\vec{\eta}) \\ &= \mathbf{D}\vec{\mu} - \nabla_{\vec{\mu}} \mathcal{L}(\vec{\mu} \mid \vec{s},\vec{a}) = \mathbf{D}\vec{\mu} - \nabla_{\vec{\mu}} \mathcal{L}(\vec{s},\vec{a},\vec{\mu})\end{aligned} \tag{12}$$

---

[9] A 'particular partition' here is not meant in the sense of a partition function in statistical physics, rather as an individuation between the states of a particle and the states external to it. However, it is interesting to note that the free energy functional—we will arrive at by partitioning external from particular states—is itself effectively a (log) partition function in the statistical physics sense.

[10] Noting that if two subsets of trajectories were independent, as opposed to being conditionally independent, we would be describing two separate systems.

[11] Equation (12) is tautologically equivalent to $\dot{\vec{\mu}}(\tau) = \mathbf{D}\vec{\mu}$. This implicit augmentation is important as it will enable to relate these dynamics to generalised Bayesian filters.

Because internal paths depend only on blanket paths there is a deterministic mapping from every blanket path to the corresponding internal path of least action. If this map is injective, this implies a map between the internal path of least action and the conditional density over external paths, given blanket paths.[12] This injection can be expressed as a *variational density*:

$$q_{\vec{\mu}}(\vec{\eta}) \triangleq p(\vec{\eta} \,|\, \vec{s}, \vec{a})$$
$$\vec{\mu} \triangleq \arg\min_{\vec{\mu}} \mathcal{L}(\vec{\mu} \,|\, \vec{s}, \vec{a}) \Leftrightarrow \nabla_{\vec{\mu}} \mathcal{L}(\vec{\mu} \,|\, \vec{s}, \vec{a}) = 0 \tag{13}$$

Equation (13) can be interpreted as saying that the internal path of least action, for any given blanket path, encodes beliefs about external paths. This licences the following definition:

**Definition 1**: *a particle is defined by a particular partition with a nonempty set of internal states, whose paths of least action parameterise a conditional density over external paths. In other words, a particular partition* (10) *defines a particle when* (13) *holds*.

Although based upon the same assumptions, the free energy principle can be seen as complementing quantum, statistical and classical mechanics with a Bayesian mechanics, by paying careful attention to the separation of external and internal dynamics (Friston, 2019). This separation is via the conditional independence in (12). By construction, this endows particles with an autonomy in the sense that active and internal (i.e., autonomous) paths, $\vec{\alpha} = (\vec{a}, \vec{\mu}) \subset \vec{x}$ do not depend on external paths, given sensory paths. The free energy principle concerns the nature of this autonomous dynamics.

**Lemma** (*variational free energy*): Consider a *variational density* over external paths parameterised by internal paths $q_{\vec{\mu}}(\vec{\eta})$, where the path of least action encodes the posterior density over external paths: $q_{\vec{\mu}}(\vec{\eta}) = p(\vec{\eta} \,|\, \vec{s}, \vec{a})$. Then internal paths of least action of particles minimise a free energy functional of Bayesian beliefs[13] about external paths:

---

[12] In the case of linear flows in (1), the flow of generalised states in (2) is also linear, which means that the Lagrangian is quadratic, e.g., see Section 6 in (Friston et al. 2022). In this case, the condition under which a variational density can be defined that satisfies (13) is the object of Lemma 2.1 in (Da Costa et al. 2021). In the case of nonlinear flows—since the space of continuous paths is an infinite-dimensional Banach space—we know that many maps between two such spaces are injective. This motivates why, for each system that could be modelled, out of all the maps one could obtain, many of them will afford (13).

[13] Beliefs here are meant in a technical, Bayesian sense; that is, as probability distributions over external variables. In this sense the free energy is a functional (i.e., function of a function) of beliefs, or equivalently, a function of their parameters.

$$\vec{\mu} = \arg\min_{\vec{\mu}} \mathcal{L}(\vec{\mu} \mid \vec{s}, \vec{a}) \Leftrightarrow \nabla_{\vec{\mu}} \mathcal{L}(\vec{s}, \vec{a}, \vec{\mu}) = 0 \Leftrightarrow$$
$$\vec{\mu} = \arg\min_{\vec{\mu}} F(\vec{s}, \vec{a}, \vec{\mu}) \Leftrightarrow \nabla_{\vec{\mu}} F(\vec{s}, \vec{a}, \vec{\mu}) = 0 \Rightarrow$$
$$\dot{\vec{\mu}} = \mathbf{D}\vec{\mu} - \nabla_{\vec{\mu}} F(\vec{s}, \vec{a}, \vec{\mu})$$

$$\begin{aligned}
F(\vec{s}, \vec{\alpha}) &= \underbrace{\mathbb{E}_{q(\vec{\eta})}[\mathcal{L}(\vec{\eta}, \vec{s}, \vec{\alpha}) + \ln q(\vec{\eta})]}_{\text{Variational free energy}} \\
&= \underbrace{\mathbb{E}_q[\mathcal{L}(\vec{s}, \vec{\alpha} \mid \vec{\eta}) + \mathcal{L}(\vec{\eta})]}_{\text{Energy constraint}} - \underbrace{\mathbb{E}_q[-\ln q(\vec{\eta})]}_{\text{Entropy}} \\
&= \underbrace{D_{KL}[q(\vec{\eta}) \mid p(\vec{\eta})]}_{\text{Complexity}} + \underbrace{\mathbb{E}_q[\mathcal{L}(\vec{s}, \vec{\alpha} \mid \vec{\eta})]}_{(-)\text{ Accuracy}} \\
&= \underbrace{D_{KL}[q(\vec{\eta}) \| p(\vec{\eta} \mid \vec{s}, \vec{a})]}_{\text{Divergence}} + \underbrace{\mathcal{L}(\vec{s}, \vec{\alpha})}_{(-)\text{ Log evidence}} \geq \mathcal{L}(\vec{s}, \vec{\alpha})
\end{aligned} \quad (14)$$

This variational free energy can be rearranged in several ways. First, it can be expressed as an energy constraint minus an entropy, which licences the name free energy (Feynman, 1972). In this decomposition, minimising variational free energy corresponds to the maximum entropy principle, under the constraint that the energy is minimised (Jaynes, 1957; Lasota and Mackey, 1994; Sakthivadivel, 2022b). The energy constraint is a functional of the marginal density over external and particular paths, which plays the role of a *generative model*; namely, a joint density over causes (i.e., external paths) and their consequences (i.e., particular paths). In (14) the generative model has been expressed in terms of a *likelihood* and *prior* Lagrangian that can be read as specifying characteristic or preferred trajectories in the space of causes (i.e., external states) or consequences (i.e., particular states).

Second—on a statistical reading—variational free energy can be decomposed into the (negative) log likelihood of particular paths (i.e., *accuracy*) and the Kullback Leibler (KL) divergence between posterior and prior densities over external paths (i.e., *complexity*). Finally, it can be written as the negative *log evidence* plus the KL divergence between the variational and conditional (i.e., posterior) density. In variational Bayesian inference[14] (Beal, 2003), negative free energy is called an *evidence lower bound* or ELBO (Bishop, 2006; Winn and Bishop, 2005). It is called a bound because the KL divergence is never less than zero.

**Proof**: recall that the Lagrangian is a surprisal, and the KL divergence between two densities is the expected difference in their surprisals. The various decompositions of the variational free energy follows from this observation. The proof is straightforward by construction: substituting the definition of the variational density in (13) into the final equality in (14) shows that the Lagrangian and variational free energy of particular paths share the same minima:

---
[14] When the set of active states is empty, and the particle is inert.

$$F(\vec{s},\vec{a},\vec{\mu}) = \underbrace{D_{KL}[q_{\vec{\mu}}(\vec{\eta}) \| p(\vec{\eta}|\vec{s},a)]}_{=0} + \mathcal{L}(\vec{s},\vec{a},\vec{\mu}) = \mathcal{L}(\vec{s},\vec{a},\vec{\mu})$$

$$\Rightarrow$$

$$\dot{\vec{\mu}}(\tau) = \mathbf{D}\vec{\mu} - \nabla_{\tilde{\mu}}\mathcal{L}(\vec{s},\vec{a},\vec{\mu})$$

$$= \mathbf{D}\vec{\mu} - \nabla_{\tilde{\mu}}F(\vec{s},\vec{a},\vec{\mu})$$

(15)

Replacing the Lagrangian in (12) with variational free energy gives (14).

**Corollary**: if autonomous and external paths are conditionally independent, given sensory paths, then the autonomous path of least action can also be cast as a gradient flow on variational free energy

$$(\vec{\eta} \perp \vec{\alpha})|\vec{s} \Rightarrow \dot{\vec{\alpha}} = \mathbf{D}\vec{\alpha} - \nabla_{\tilde{\alpha}}F(\vec{s},\vec{\alpha}) \tag{16}$$

Where $\vec{\alpha} \triangleq \arg\min_{\tilde{\alpha}} \mathcal{L}(\vec{\alpha}|\vec{s})$. In this case, the most likely paths of *both internal and active states* perform a gradient descent on variational free energy.

**Proof**: the conditional independence above means that the autonomous path of least action is a gradient flow on the Lagrangian of particular states:

$$(\vec{\eta} \perp \vec{\alpha})|\vec{s} \Leftrightarrow \mathcal{L}(\vec{\alpha}|\vec{s},\vec{\eta}) = \mathcal{L}(\vec{\alpha}|\vec{s})$$

$$\Rightarrow$$

$$\dot{\vec{\alpha}} = \mathbf{D}\vec{\alpha} - \nabla_{\tilde{\alpha}}\mathcal{L}(\vec{\alpha}|\vec{s},\vec{\eta}) = \mathbf{D}\vec{\alpha} - \nabla_{\tilde{\alpha}}\mathcal{L}(\vec{\alpha}|\vec{s}) = \mathbf{D}\vec{\alpha} - \nabla_{\tilde{\alpha}}\mathcal{L}(\vec{s},\vec{\alpha})$$

$$= \mathbf{D}\vec{\alpha} - \nabla_{\tilde{\alpha}}F(\vec{s},\vec{\alpha})$$

(17)

Where the motion of autonomous paths of least action follows from **Error! Reference source not found.** and the Lagrangian of particular states is the variational free energy of internal paths of least action by (15)

**Remark**: The functional form of variational free energy licences a representational interpretation of internal dynamics: the internal paths of least action play the role of sufficient statistics or parameters of Bayesian beliefs about external dynamics. On this view, a gradient flow on variational free energy corresponds to minimising the complexity of Bayesian beliefs about external states, while providing accurate predictions of the dynamics of a particle's sensory (and autonomous) states. Minimising complexity means that internal paths of least action encode Bayesian beliefs about external paths that are as close as possible to prior beliefs. The paths of least action can either be interpreted as the behaviour of a particle with classical mechanics (see below), in which random fluctuations are negligible. Conversely, they can be considered as the average responses over multiple realisations of the same dynamics, of the sort solicited by event related averaging in various fields; e.g., (Licata and Chiatti, 2019; Singh et al., 2002). In this case, we can think of the paths of least action as expressing the average (e.g., ensemble) behaviour of a system.

The variational free energy lemma says that the internal paths of least action minimise a variational free energy functional. So, is minimisation of variational free energy an explanation for—or a description of—internal dynamics? The question is itself revealing. It is a question about internal dynamics. However, the internal dynamics are unknowable (unmeasurable) by construction. Only blanket states are observable (measurable). This means one can only say that it *looks as if* internal dynamics are inferring external paths, under certain conditions: the existence of a variational density.

# A typology of particular kinds

We now begin to unpack a typology of particular kinds. First, we distinguish between active and inert particles, which do and do not have active states. We will then distinguish conservative and dissipative particles, for which random fluctuations can be, and cannot be, ignored. In the final sections, we distinguish between ordinary and strange particles, whose active states do, and do not, directly influence internal states. See Figure 2.

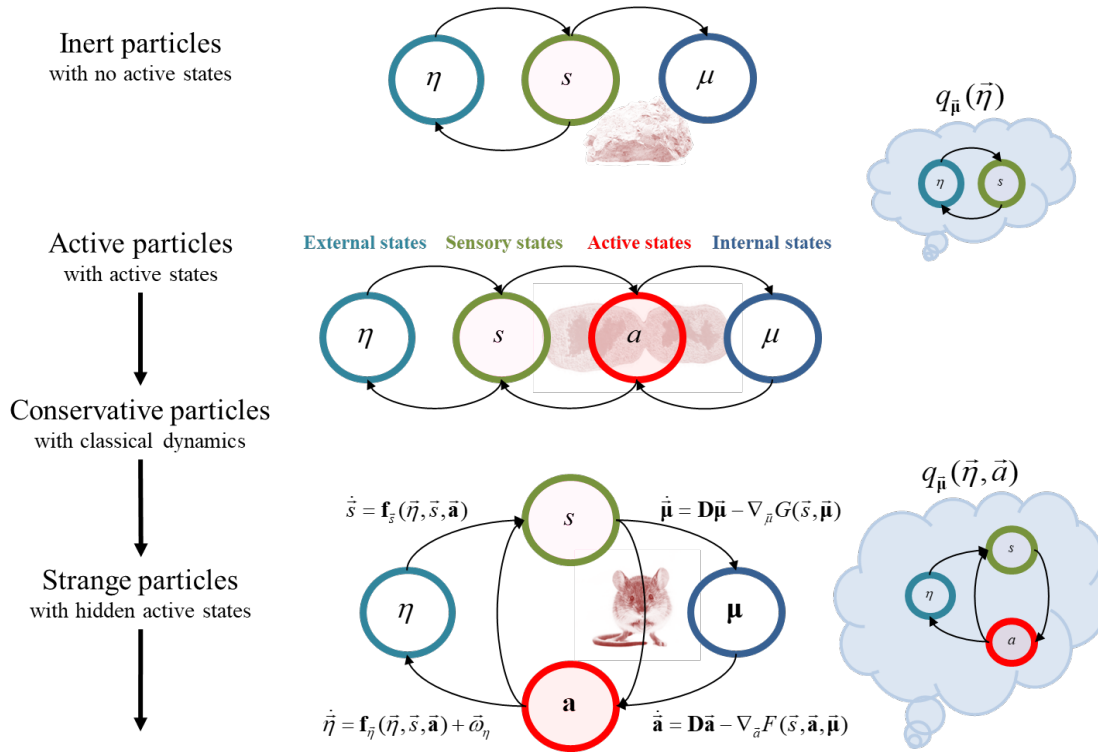

**Figure 2** – *particular kinds*. This schematic illustrates the definitive features of the particular kinds considered in the main text. The upper panels show inert and active particles, without and with active states, respectively. When the particular states of an active particle follow paths of least action, we have conservative particles. When the active states of conservative particles are hidden from (i.e., do not directly influence) internal states, we have strange particles. Strange particles will appear to infer their own actions in virtue of behaving as if their internal dynamics encode a variational density (i.e., Bayesian belief) over the hidden causes of sensations; namely external *and active paths*.

The inferential process of internal states can only manifest to the external world when internal dynamics influence blanket flows, which, by construction, require active states (noting that internal states cannot influence sensory states). This speaks to two kinds of particles with and without active states: namely, active and inert particles, respectively.

**Definition 2**: *an active particle is a particle with a nonempty set of active states. Conversely, an inert particle is a particle with no active states.*

This definition precludes any form of panpsychism when applying the free energy principle, in the sense that any ontic or noetic claims can only concern active particles, or active matter (Ramaswamy, 2010). Put simply, we can never observe internal dynamics directly because these are hidden behind the Markov blanket. Whether internal paths of least action parameterise beliefs about external paths and therefore minimise variational free energy can only manifest via active states; that is, in active particles. Since these are non-existent in inert particles, there is no way to observe whether a given inert particle minimises variational free energy (e.g., the interior of a black hole). Before turning to the behaviour of active particles, we consider inert particles through the lens of inference.

## Inert particles and Bayesian filtering

Although internal dynamics are hidden behind their Markov blanket, we can simulate or emulate (the most likely behaviour of) *inert* particles by integrating the following equations of generalised motion, inserting (14) into (10):

$$\begin{bmatrix} \dot{\vec{\eta}} \\ \dot{\vec{s}} \\ \dot{\vec{\mu}} \end{bmatrix} = \begin{bmatrix} \mathbf{f}_\eta(\vec{\eta},\vec{s}) + \vec{\omega}_\eta \\ \mathbf{f}_s(\vec{\eta},\vec{s}) + \vec{\omega}_s \\ \mathbf{D}\vec{\mu} - \nabla_{\vec{\mu}} F(\vec{s},\vec{a},\vec{\mu}) \end{bmatrix} \qquad (18)$$

Here, internal dynamics are expressed as a generalised gradient flow on variational free energy, while external and sensory paths are specified with equations of motion and the statistics of random fluctuations. In statistics, these constitute a *state-space model*. The functional form of variational free energy can be unpacked under the Laplace approximation (Friston et al., 2007; Ramstead et al., 2022). For example, if $\sigma(\vec{\mu}) = \arg\max q_{\vec{\mu}}(\vec{\eta}) = \vec{\eta}$—the map from the internal to the external path of least action—is well-defined, we have[15]

---

[15] Omitting constant terms for clarity.

$$F(\vec{s},\vec{\mu}) = \underbrace{\mathbb{E}_{\mathbf{q}}[\mathcal{L}(\vec{\eta},\vec{s}) + \ln q(\vec{\eta})]}_{\text{Variational free energy}}$$

$$\approx \underbrace{\mathcal{L}(\vec{s} \mid \sigma(\vec{\mu}))}_{\text{Likelihood constraint}} + \underbrace{\mathcal{L}(\sigma(\vec{\mu}))}_{\text{Prior constraint}} - \underbrace{\tfrac{1}{2}\ln |\Sigma(\vec{\mu})|}_{\text{Entropy}} + \underbrace{\tfrac{1}{2} tr(\Sigma(\vec{\mu})\nabla^2 \mathcal{L})}_{\text{constant (dim }\vec{\mu})}$$

$$q(\vec{\eta}) = \mathcal{N}(\vec{\eta} : \sigma(\vec{\mu}), \Sigma(\vec{\mu})), \quad \Sigma(\vec{\mu})^{-1} = \nabla^2 \mathcal{L} \tag{19}$$

$$\mathcal{L}(\vec{\eta},\vec{s}) = \vec{\omega}_\eta \cdot \tfrac{1}{4\Gamma_\eta} \vec{\omega}_\eta + \vec{\omega}_s \cdot \tfrac{1}{4\Gamma_s} \vec{\omega}_s + \mathcal{S}_\eta + \mathcal{S}_s$$

$$\vec{\omega}_\eta = \mathbf{D}\vec{\eta} - \mathbf{f}_\eta(\vec{\eta},\vec{s})$$

$$\vec{\omega}_s = \mathbf{D}\vec{s} - \mathbf{f}_s(\vec{\eta},\vec{s})$$

This functional form renders the internal dynamics in (18) equivalent to a generalised Bayesian filter (Friston et al., 2010b); namely, the generalisation of an extended Kalman-Bucy filter (Loeliger, 2002; Schiff and Sauer, 2008) to high order motion. In this setting, internal states parameterise a Gaussian posterior over external states—a.k.a., variational Laplace (Friston et al., 2007).

## Conservative particles and expected free energy

"*It is implied that, in some sense, a rudimentary consciousness is present even at the level of particle physics*" (Bohm, 2008).

We now turn to the distinction between microscopic and macroscopic particles, where macroscopic particles possess classical mechanics, in which random fluctuations are (almost) averaged away[16] and the trajectories of particular states are (almost) paths of least action. This introduces the next key distinction; namely, between *dissipative* and *conservative* particles whose particular dynamics are, and are not, subject to random fluctuations, respectively.

**Definition 3**: *a conservative (classical) particle is an active particle whose particular states follow paths of least action. Equivalently, it is an active particle whose random fluctuations on particular states have amplitudes that are infinitesimally small. In other words, we consider the limiting regime where the covariance of the random fluctuations $\omega_s(\tau), \omega_a(\tau), \omega_\mu(\tau)$ in (10) tends to zero $\Gamma_\pi \to 0$.*

Conservative (or classical) particles are especially interesting because they respond deterministically and systematically to external influences, in virtue of the fact random fluctuations on their dynamics (almost) disappear. It could be argued that all macroscopic particles are conservative, simply because their random fluctuations are averaged away (Friston, 2019). This means that their dynamics rests on chaotic itinerancy that is underwritten by nonlinear coupling, breaking of detailed balance, and associated nonequilibria (Friston et al., 2021; Lasota and Mackey, 1994; Da Costa and Pavliotis, 2022). Removing uncertainty about particular paths has an important consequence. Because there is no uncertainty about autonomous paths, given sensory paths, there is no further information about external paths afforded by autonomous paths, rendering them conditionally independent. This can be expressed in terms of expected Lagrangians using a bound argument. From (9):

$$\begin{aligned}
&\mathbb{E}[\mathcal{L}(\vec{\alpha} \mid \vec{s})] \geq \mathbb{E}[\mathcal{L}(\vec{\alpha} \mid \vec{s}, \vec{\eta})] \geq \mathcal{S}_0 \\
&\lim_{\Gamma_\pi \to 0} \mathbb{E}[\mathcal{L}(\vec{\alpha} \mid \vec{s})] = \mathbb{E}[\mathcal{L}(\vec{\alpha} \mid \vec{s}, \vec{\eta})] = \mathcal{S}_0 \\
&\Rightarrow \mathbb{E}_{p(\vec{s}, \vec{\eta})}[D_{KL}[p(\vec{\alpha} \mid \vec{s}, \vec{\eta}) \parallel p(\vec{\alpha} \mid \vec{s})]] = 0 \\
&\Rightarrow \mathcal{L}(\vec{\alpha} \mid \vec{s}, \vec{\eta}) = \mathcal{L}(\vec{\alpha} \mid \vec{s}) \Leftrightarrow (\vec{\alpha} \perp \vec{\eta}) \mid \vec{s} \\
&\Rightarrow \mathcal{L}(\vec{\eta} \mid \vec{s}, \vec{\alpha}) = \mathcal{L}(\vec{\eta} \mid \vec{s}) \Leftrightarrow (\vec{\eta} \perp \vec{\alpha}) \mid \vec{s}
\end{aligned} \qquad (20)$$

---

[16] Physically, this means that we are considering a description of particles at a sufficiently high-level of coarse graining so that microscopic fluctuations are almost averaged away, and the laws of classical mechanics prevail. See (Friston et al. 2020) for examples using the coarse graining apparatus of the renormalisation group.

In other words, although active states influence external states, the active states are determined completely by sensory paths, as they provide a Markov blanket separating external and autonomous states. Please see Figure 3 for a diagrammatic illustration.

Crucially, this conditional independence means that conservative particles satisfy (16), which yields an expression for the most likely behaviour of conservative particles

$$\begin{bmatrix} \dot{\vec{\eta}} \\ \dot{\vec{s}} \\ \dot{\vec{a}} \\ \dot{\vec{\mu}} \end{bmatrix} = \begin{bmatrix} \mathbf{f}_\eta(\vec{\eta}, \vec{s}, \vec{a}) + \vec{\omega}_\eta \\ \mathbf{f}_s(\vec{\eta}, \vec{s}, \vec{a}) + \vec{\omega}_s \\ \mathbf{D}\vec{a} - \nabla_{\vec{a}} F(\vec{s}, \vec{a}, \vec{\mu}) \\ \mathbf{D}\vec{\mu} - \nabla_{\vec{\mu}} F(\vec{s}, \vec{a}, \vec{\mu}) \end{bmatrix}. \qquad (21)$$

These equations of motion allow one to simulate the behaviour of conservative particles as a (generalised) gradient flow of autonomous states on variational free energy. The resulting dynamics can be read as a generalised homeostasis (cf., the behaviour of a thermostat or noise cancellation scheme) or, in control theory, control as inference (Baltieri and Buckley, 2019; Kappen, 2005; Todorov, 2008). In short, active states will look as if they are trying to minimise the sensory prediction errors in (19). More generally, the active paths of conservative particles will look as if they are trying to maximise the *accuracy* part of variational free energy, thereby fulfilling the predictions encoded by internal dynamics (because the *complexity* part does not depend upon active paths). This interpretation furnishes an elementary but expressive formulation of active perception or inference (Friston et al., 2010a).

Can we say anything more definitive about the ensuing autonomous behaviour? We will see that the autonomous dynamics of conservative particles acquire a purposeful aspect—a purpose that can be articulated in terms of information and preference seeking behaviour with the following lemma (Barp et al. 2022; Friston et al. 2022):

**Lemma** (*expected free energy*): the Lagrangian of autonomous paths of conservative particles can be expressed as a free energy functional that entails optimal Bayesian design and decision-making:

$$E(\vec{\alpha}) = \underbrace{\mathbb{E}_{p(\vec{\eta},\vec{s}|\vec{\alpha})}[\mathcal{L}(\vec{\eta},\vec{s}) - \mathcal{L}(\vec{\eta}\,|\,\vec{\alpha})]}_{\text{Expected free energy}} = \mathcal{L}(\vec{\alpha})$$

$$= \underbrace{D_{KL}[p(\vec{s}\,|\,\vec{\alpha})\,\|\,p(\vec{s})]}_{\text{Risk}} + \underbrace{\mathbb{E}_{p(\vec{\eta},\vec{s}|\vec{\alpha})}[\mathcal{L}(\vec{s}\,|\,\vec{\eta},\vec{\alpha})]}_{\text{Ambiguity} = \mathcal{S}_0} \quad (22)$$

$$= \underbrace{\mathbb{E}_{p(\vec{s}|\vec{\alpha})}[\mathcal{L}(\vec{s})]}_{\text{Expected cost}} - \underbrace{\mathbb{E}_{p(\vec{s}|\vec{\alpha})}[D_{KL}[p(\vec{\eta}\,|\,\vec{s},\vec{\alpha})\,\|\,p(\vec{\eta}\,|\,\vec{\alpha})]]}_{\text{Expected information gain}}$$

The functional forms of variational (14) and expected free energy (22) suggest that expected (negative) *accuracy* becomes *ambiguity*, while expected *complexity* becomes *risk*. Similarly, the expected (negative) *divergence* becomes *expected information gain*. If we read the Lagrangian as a cost function of sensory outcomes, the most likely autonomous paths of conservative particles will look as if they are maximising *expected information gain*, while minimising *expected cost*. These expectations underwrite the principles of optimal Bayesian design and decision theory, respectively.

**Proof**: when random fluctuations on the motion of particular states vanish, there is no uncertainty about their paths given their parents. This means there is no uncertainty about particular paths, given external paths; no uncertainty about autonomous paths, given sensory paths and no uncertainty about sensory paths, given external and autonomous paths. Expressed in terms of continuous and differential entropies this means, from (9):

$$\begin{aligned}
\lim_{\Gamma_\pi \to 0} \mathbb{H}[p(\vec{\pi}\,|\,\vec{\eta})] &= 0 \Leftrightarrow \mathbb{E}[\mathcal{L}(\vec{\pi}\,|\,\vec{\eta})] = \mathcal{S}_0 \\
\mathbb{H}[p(\vec{\alpha}\,|\,\vec{s})] &= 0 \Leftrightarrow \mathbb{E}[\mathcal{L}(\vec{\alpha}\,|\,\vec{s})] = \mathcal{S}_0 \Rightarrow \\
\mathbb{H}[p(\vec{s}\,|\,\vec{\eta},\vec{\alpha})] &= 0 \Leftrightarrow \mathbb{E}[\mathcal{L}(\vec{s}\,|\,\vec{\eta},\vec{\alpha})] = \mathcal{S}_0 \Rightarrow \mathbb{E}_{p(\vec{\eta},\vec{s}|\vec{\alpha})}[\mathcal{L}(\vec{s}\,|\,\vec{\eta},\vec{\alpha})] = \mathcal{S}_0 \\
\mathbb{H}[p(\vec{\alpha}\,|\,\vec{\eta},\vec{s})] &= 0 \Leftrightarrow \mathbb{E}[\mathcal{L}(\vec{\alpha}\,|\,\vec{\eta},\vec{s})] = \mathcal{S}_0 \Rightarrow \mathbb{E}_{p(\vec{\eta},\vec{s}|\vec{\alpha})}[\mathcal{L}(\vec{\alpha}\,|\,\vec{\eta},\vec{s})] = \mathcal{S}_0
\end{aligned} \quad (23)$$

These constraints allow us to express the Lagrangian of autonomous paths as an expected free energy. From (23), we have:

$$\begin{aligned}
\mathcal{L}(\vec{\alpha}) &= \mathbb{E}_{p(\vec{\eta},\vec{s}|\vec{\alpha})}[\mathcal{L}(\vec{\eta},\vec{s},\vec{\alpha}) - \mathcal{L}(\vec{\eta},\vec{s}\,|\,\vec{\alpha})] \\
&= \mathbb{E}_{p(\vec{\eta},\vec{s}|\vec{\alpha})}[\mathcal{L}(\vec{\eta},\vec{s}) - \mathcal{L}(\vec{\eta}\,|\,\vec{\alpha})] + \mathbb{E}_{p(\vec{\eta},\vec{s}|\vec{\alpha})}[\mathcal{L}(\vec{\alpha}\,|\,\vec{\eta},\vec{s}) - \mathcal{L}(\vec{s}\,|\,\vec{\eta},\vec{\alpha})] \\
&= \underbrace{\mathbb{E}_{p(\vec{\eta},\vec{s}|\vec{\alpha})}[\mathcal{L}(\vec{\eta},\vec{s}) - \mathcal{L}(\vec{\eta}\,|\,\vec{\alpha})]}_{\text{Expected free energy}} + \mathcal{S}_0 - \mathcal{S}_0 = E(\vec{\alpha})
\end{aligned} \quad (24)$$

Substituting the final equality in (20) into (24) gives (22):

$$E(\vec{\alpha}) = \underbrace{\mathbb{E}_{p(\vec{\eta},\vec{s}|\vec{\alpha})}[\mathcal{L}(\vec{s}) + \mathcal{L}(\vec{\eta}\,|\,\vec{s},\vec{\alpha}) - \mathcal{L}(\vec{\eta}\,|\,\vec{\alpha})]}_{\text{Expected free energy}}$$

$$= \underbrace{\mathbb{E}_{p(\vec{s}|\vec{\alpha})}[\mathcal{L}(\vec{s})]}_{\text{Expected cost}} - \underbrace{\mathbb{E}_{p(\vec{s}|\vec{\alpha})}[D_{KL}[p(\vec{\eta}\,|\,\vec{s},\vec{\alpha}) \,\|\, p(\vec{\eta}\,|\,\vec{\alpha})]]}_{\text{Expected information gain}} \quad (25)$$

$$= \underbrace{\mathbb{E}_{p(\vec{s}|\vec{\alpha})}[\mathcal{L}(\vec{s})]}_{\text{Expected cost}} - \underbrace{\mathbb{E}_{p(\vec{\eta}|\vec{\alpha})}[D_{KL}[p(\vec{s}\,|\,\vec{\eta},\vec{\alpha}) \,\|\, p(\vec{s}\,|\,\vec{\alpha})]]}_{\text{Expected information gain}}$$

$$= \underbrace{D_{KL}[p(\vec{s}\,|\,\vec{\alpha}) \,\|\, p(\vec{s})]}_{\text{Risk}} + \underbrace{\mathbb{E}_{p(\vec{\eta},\vec{s}|\vec{\alpha})}[\mathcal{L}(\vec{s}\,|\,\vec{\eta},\vec{\alpha})]}_{\text{Ambiguity} = \mathcal{S}_0}$$

These equalities rest on the fact that the KL divergences in the expected information gains are the same; i.e., the mutual information between external and sensory paths, given an autonomous path [p. 273 in (Batina et al. 2011)].

**Remarks**: the relationship among the various terms contributions to expected free energy can be seen clearly in terms of their expectations; namely, entropies. From (22):

$$\mathbb{E}_{p(\vec{\alpha})}[\mathcal{L}(\vec{\alpha})] - \mathcal{S}_0 = \mathbb{H}[p(\vec{\alpha})] = \underbrace{\mathbb{H}[p(\vec{\eta},\vec{s})] - \mathbb{H}[p(\vec{\eta}\,|\,\vec{\alpha})]}_{\text{<Expected free energy>}}$$

$$= \underbrace{\mathbb{H}[p(\vec{\eta})] - \mathbb{H}[p(\vec{\eta}\,|\,\vec{\alpha})]}_{\text{<Risk>}} \quad (26)$$

$$= \underbrace{\mathbb{H}[p(\vec{s})]}_{\text{<Cost>}} - \underbrace{\mathbb{H}[p(\vec{s}\,|\,\vec{\alpha})]}_{\text{<Information gain>}}$$
$$\underbrace{\phantom{\mathbb{H}[p(\vec{s})] - \mathbb{H}[p(\vec{s}\,|\,\vec{\alpha})]}}_{\text{<Risk>}}$$

In the case of conservative particles, the conditional uncertainty about sensory paths, given autonomous paths, is the information gain about external paths, afforded by sensory paths:

$$\mathbb{H}[p(\vec{s}\,|\,\vec{\alpha})] = \underbrace{\mathbb{H}[p(\vec{\eta}\,|\,\vec{\alpha})] - \mathbb{H}[p(\vec{\eta}\,|\,\vec{s},\vec{\alpha})]}_{\text{<Information gain>}} \quad (27)$$

Figure 3 illustrates these decompositions for people who find it easier to think in terms in terms of information diagrams.

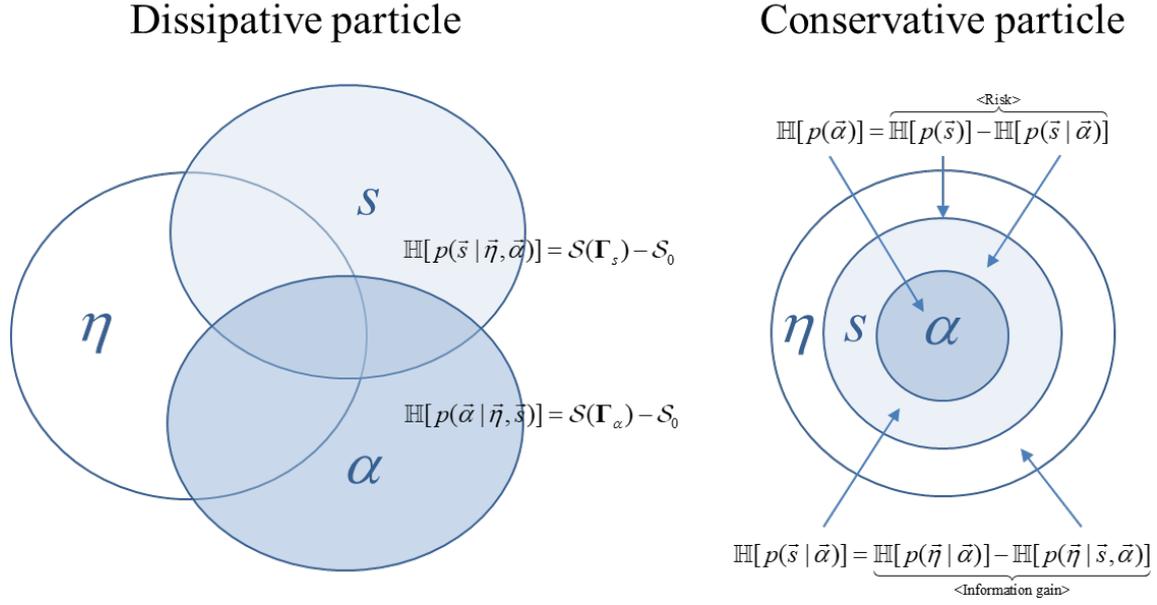

**Figure 3**: *dissipative and conservative particles*. These information diagrams depict the entropy of external, sensory and autonomous paths, where intersections correspond to shared or mutual information. A conditional entropy corresponds to an area that is outside the variable upon which it is conditioned. The diagram on the left shows the generic (dissipative) case, in which uncertainty about paths inherits from random fluctuations. When the random fluctuations on the motion of particular states vanish, we have conservative particles, in which there is (almost) no uncertainty about autonomous paths, given sensory paths and no uncertainty about particular paths, given external paths (the right information diagram). This illustrates the various ways of expressing the entropy of autonomous paths (upper equation)—or the entropy of sensory paths given autonomous paths (lower equation)—in terms of conditional entropies. See main text for a fuller description.

In summary, the Lagrangian of autonomous paths can be expressed as a divergence; namely, the divergence between the density over sensory (or external) paths with and without autonomous paths. This can be regarded as the objective function used in engineering to optimise the trajectory of control variables in *model predictive control* (Schwenzer et al., 2021). An equivalent neurobiological perspective is provided by *perceptual control theory*, in which sensory states are maintained near a preferred value through the consequences of action (Mansell, 2011). On this view, risk is the divergence between sensory paths, given an autonomous path, and the marginal density over sensory paths. This marginal density encodes preferences, because they constitute the priors of the generative model, over the sensory consequences of action (Parr and Friston, 2019). Intuitively, these paths can be read as 'outcomes', 'events' or 'narratives' the particle expects to encounter.

Expected free energy supplements risk with ambiguity; namely, the expected inaccuracy or negative sensory likelihood, given an autonomous path. Clearly, this is redundant as a description of conservative particles that—by definition—have minimal ambiguity (because there are no random sensory fluctuations). However, in applications of the FEP, the ambiguity term is retained to ensure the particle or agent seeks out unambiguous regimes of state space; thereby, evincing the behaviour of conservative particles.

In discrete state-space formulations, ambiguity is usually expressed as the expected conditional uncertainty (i.e., entropy) of sensory states, given external states by assuming conditional independence between sensory and external states. Under this assumption we have, from (20):

$$\mathbb{E}_{p(\vec{\eta},\vec{s}|\vec{\alpha})}[\mathcal{L}(\vec{s}\,|\,\vec{\eta})] = \underbrace{\mathbb{E}_{p(\vec{\eta}|\vec{\alpha})}[\mathbb{E}_{p(\vec{s}|\vec{\eta})}[\mathcal{L}(\vec{s}\,|\,\vec{\eta})]]}_{\text{Ambiguity}} = \underbrace{\mathbb{E}_{p(\vec{s}|\vec{\alpha})}[\mathbb{E}_{p(\vec{\eta}|\vec{s})}[\mathcal{L}(\vec{s}\,|\,\vec{\eta})]]}_{\text{Expected inaccuracy}} \tag{28}$$

This is why the expected inaccuracy is referred to as ambiguity. In brief, the expected free energy lemma, suggests that conservative particles can be described as minimising the risk of incurring external trajectories that diverge from prior preferences, while avoiding ambiguous states of affairs.

The final term in (22) provides a Bayesian interpretation of expected free energy. It is the expected divergence between posterior beliefs about external paths, given autonomous paths, with and without sensory paths. In other words, it scores the resolution of uncertainty or expected information gain[17] afforded by sensory outcomes. This is the objective function in optimal Bayesian design and active learning (Lindley, 1956; Mackay, 1992). In this sense, it is sometimes referred to as intrinsic value or epistemic affordance (Friston et al., 2017c). The remaining term is the expected Lagrangian, which scores the probability that a particle will sample preferred sensory trajectories. In the setting of Bayesian decision theory, this minimises expected cost (Berger, 2011). In this sense it is sometimes referred to as extrinsic value or pragmatic affordance (Friston et al., 2017a; Schwartenbeck et al., 2015). Note that expected free energy dissolves the exploration-exploitation dilemma (Cohen et al., 2007), because there is a unique course or path (of least) action that subsumes explorative, information-seeking and exploitative, preference-seeking imperatives. Please see Figure 4 for a schematic that links and contextualises these normative accounts of behaviour through expected free energy. On this

---

[17] Expected information gain can be expressed in a number of ways. The expression in (23) is perhaps the most intuitive, showing the degree to which the conditional density of external paths changes once we take sensory paths into account (i.e., how far we update our beliefs about external paths given sensory paths). It can also be expressed as a mutual information, which scores the degree of conditional dependence between sensory and external paths, or as the difference between two entropies:

$$\underbrace{D_{KL}[p(\vec{\eta},\vec{s}\,|\,\vec{\alpha})\,\|\,p(\vec{s}\,|\,\vec{\alpha})p(\vec{\eta}\,|\,\vec{\alpha})]}_{\text{Mutual information}} = \underbrace{\mathbb{E}_{p(\vec{s}|\vec{\alpha})}[\mathcal{L}(\vec{s}\,|\,\vec{\alpha})] - \mathbb{E}_{p(\vec{s},\vec{\eta}|\vec{\alpha})}[\mathcal{L}(\vec{s}\,|\,\vec{\eta},\vec{\alpha})]}_{\text{Expected information gain}}$$

This formulation offers an intuition as to the factors that determine the degree of belief-updating. If the entropy of sensory paths (given autonomous paths) is very large, this implies a high level of uncertainty which, if resolved, leads to a substantial change in our beliefs about external paths. The extent to which this uncertainty is resolvable depends upon the conditional entropy of sensory paths given external and autonomous paths. In other words, if there is a precise relationship between external and sensory paths, and we are uncertain about sensory paths, there must be uncertainty about external paths which is resolvable by observing sensory paths.

For conservative particles, the expected information gain reduces to the entropy of sensory paths given autonomous paths. It may therefore seem that the conservative particle assumption renders the expected information gain trivial. However, it is worth noting that it may be possible to express the dynamics of a non-conservative particle (i.e., with sensory fluctuations) as a conservative particle, simply by treating the sensory fluctuations as (fast) external states. In this case, the equality between expected information gain and the entropy of sensory paths, given autonomous paths, holds in general.

view, the FEP may provide a first principles endorsement—in terms of conservative particles—of these normative accounts.

In summary, conservative particles look as if they have purposeful behaviour; in the sense that they actively seek out preferred sensations, while trying to resolve uncertainty about the causes of those sensations. Crucially, the sensorium (i.e., external and sensory paths) will not exhibit this kind of behaviour. This is because conservative particles—in a dissipative world—break the statistical symmetry between internal and external dynamics.

The next distinction—between different kinds of things—rests on the circular causality inherent in the coupling that underwrites 'thingness'. As explained below, this causal structure depends on active paths (i.e., actions), which depend upon internal beliefs, which encode the sensory consequences of action. This causal recursion—cf., strange loop (Hofstadter, 2007) and cybernetic feedback loop (Ashby 1956)—compels us to consider an internal representation of action, which is distinct from the active states that realise those beliefs. In the final section, we consider the principles of least action that apply to certain particles that manifest a stronger kind of sentient behaviour; namely, planning.

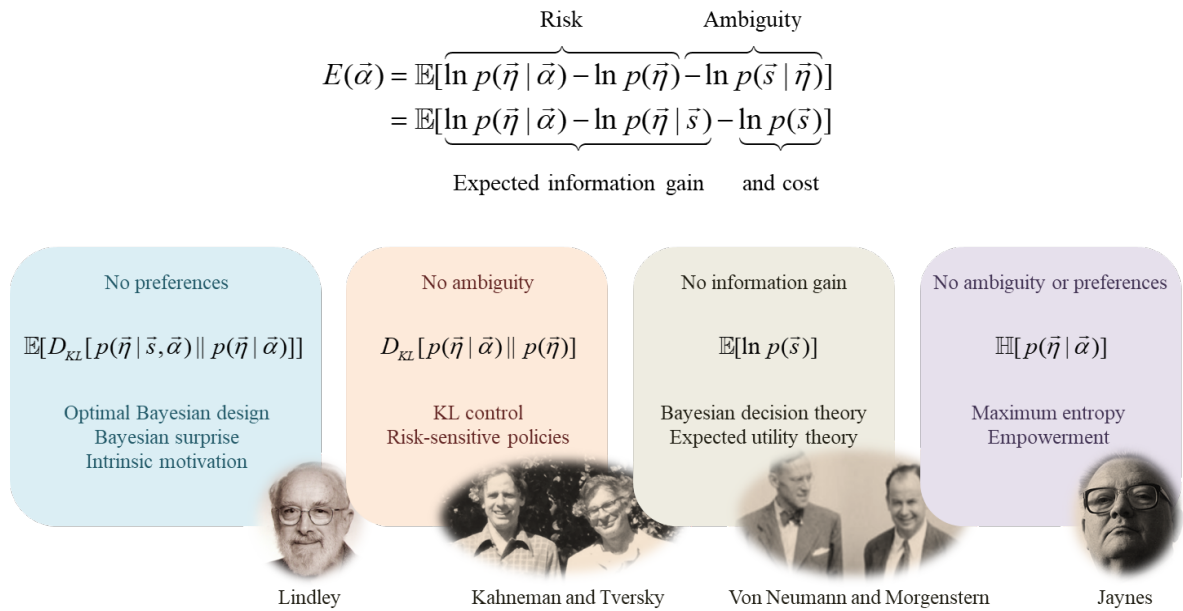

**Figure 4** – *Expected free energy and other normative accounts*. This figure illustrates the various ways in which expected free energy can be unpacked. Expected free energy subsumes several objective functions that predominate across psychology, machine learning and economics. These special cases are disclosed when one removes particular sources of uncertainty. For experimental science, this suggests which normative account is most appropriate to describe behaviour based upon the nature of uncertainty. For example, if we ignore preferences, or cost, then the expected free energy reduces to expected information gain (Lindley, 1956; MacKay, 2003) that underwrites optimum Bayesian design in statistics—or intrinsic motivation in machine learning and robotics (Barto et al., 2013; Oudeyer and Kaplan, 2007; Ryan and Deci, 1985). This is mathematically the same as expected Bayesian surprise and mutual information that underwrites salience in visual search (Itti and Baldi, 2009; Sun et al., 2011) and the computational architecture of our visual apparatus (Barlow, 1961; Barlow, 1974; Linsker, 1990; Optican and Richmond, 1987). Ignoring ambiguity leads to risk-sensitive policies in economics

(Fleming and Sheu, 2002; Kahneman and Tversky, 1979) or KL control in engineering (Todorov, 2008; van den Broek et al., 2010). Here, minimising risk corresponds to aligning predictions to prior preferences. In the absence of expected information gain (i.e., reducible uncertainty), we are left with or expected preferences or utility in economics (Von Neumann and Morgenstern, 1944)—a construction that underwrites reinforcement learning (Sutton and Barto, 1998) and behavioural psychology. Maximising expected utility under uncertainty leads to Bayesian decision theory (Berger, 2011). Finally, if we consider an unambiguous world with uninformative preferences, expected free energy reduces to the negative entropy of posterior beliefs about the causes of data. This is the maximum entropy principle, proposed as a method of inference by Jaynes (Jaynes, 1957; Lasota and Mackey, 1994). When conditioned on action, this corresponds to a form of empowerment (Klyubin et al., 2005); namely, keeping options open.

## Strange particles and generalised free energy

"*What I mean by "strange loop" is ... the cycling-around ... from one level of abstraction (or structure) to another, which feels like an upwards movement in a hierarchy, and yet somehow the successive 'upward' shifts turn out to give rise to a closed cycle. That is, despite one's sense of departing ever further from one's origin, one winds up, to one's shock, exactly where one had started out.*" (Hofstadter, 2007) *p.*101

So far, we have considered how the dynamics of conservative (i.e., classical) particles can be read as an elementary form of Bayesian inference (i.e., Bayesian mechanics). We now turn to particles that admit particles within particles. In other words, particles whose internal states have particular partitions (Palacios et al., 2020) that equip a particle with a hierarchical or deep generative model. This means there exist some internal states that are not directly influenced by active states.

We will consider the simplest instance of this sparse coupling, in which the active states of a conservative particle are hidden from internal states. In other words, internal states are only influenced by sensory states. In this kind of particle, the influence of active states on internal states must be mediated vicariously via external (or sensory) states. Figure 2 illustrates the ensuing causal architecture, where the generative model—and implicit variational density—acquire a hierarchical depth.

If internal paths are only directly influenced by sensory paths, which furnish a Markov blanket that renders the internal states conditionally independent of external and active states, this can be expressed in terms of expected Lagrangians as follows, following (20):

$$\begin{aligned}
&\mathbb{E}[\mathcal{L}(\vec{\mu} \,|\, \vec{s})] \geq \mathbb{E}[\mathcal{L}(\vec{\mu} \,|\, \vec{s}, \vec{a}, \vec{\eta})] \geq \mathcal{S}_0 \\
&\lim_{\Gamma_\pi \to 0} \mathbb{E}[\mathcal{L}(\vec{\mu} \,|\, \vec{s})] = \mathbb{E}[\mathcal{L}(\vec{\mu} \,|\, \vec{s}, \vec{a}, \vec{\eta})] = \mathcal{S}_0 \\
&\quad \Rightarrow \mathbb{E}_{p(\vec{s}, \vec{\eta}, \vec{a})}[D_{KL}[p(\vec{\mu} \,|\, \vec{s}, \vec{a}, \vec{\eta}) \,\|\, p(\vec{\mu} \,|\, \vec{s})]] = 0 \\
&\quad \Rightarrow \mathcal{L}(\vec{\mu} \,|\, \vec{s}, \vec{a}, \vec{\eta}) = \mathcal{L}(\vec{\mu} \,|\, \vec{s}) \Leftrightarrow (\vec{\mu} \perp \vec{\eta}, \vec{a}) \,|\, \vec{s} \\
&\quad \Rightarrow \mathcal{L}(\vec{\eta}, \vec{a} \,|\, \vec{s}, \vec{\mu}) = \mathcal{L}(\vec{\eta}, \vec{a} \,|\, \vec{s})
\end{aligned} \quad (29)$$

This means the internal paths of least action now become:

$$\begin{aligned}\dot{\vec{\mu}}(\tau) &= \mathbf{D}\vec{\mu} - \nabla_{\vec{\mu}}\mathcal{L}(\vec{\mu}\,|\,\vec{s},\vec{a},\vec{\eta}) = \mathbf{D}\vec{\mu} - \nabla_{\vec{\mu}}\mathcal{L}(\vec{\mu}\,|\,\vec{s}) \\ &= \mathbf{D}\vec{\mu} - \nabla_{\vec{\mu}}\mathcal{L}(\vec{s},\vec{\mu})\end{aligned} \qquad (30)$$

Because the internal paths depend only directly on sensory paths, there is a deterministic mapping from every sensory path to the corresponding internal path of least action. If this mapping is injective, this implies a map between the internal path of least action and the conditional density over external *and* active paths, given sensory paths. This means the variational density can be defined stipulatively as a Bayesian belief over the whole sensorium[18]

$$\begin{aligned} q_{\vec{\mu}}(\vec{\eta},\vec{a}) &\triangleq p(\vec{\eta},\vec{a}\,|\,\vec{s}) \\ \vec{\mu} &\triangleq \arg\min_{\vec{\mu}} \mathcal{L}(\vec{\mu}\,|\,\vec{s}) \Leftrightarrow \nabla_{\vec{\mu}}\mathcal{L}(\vec{\mu}\,|\,\vec{s}) = 0 \end{aligned} \qquad (31)$$

We will call these kinds of particles *strange*, where active states become hidden causes, lending action a certain opacity (Limanowski and Friston, 2018), and where internal paths of least action states encode beliefs about external *and* active paths. This leads to a notion of autonomy or agency, in the sense that the particle has beliefs about its own actions, and a sense of selfhood: the 'I' that follows beliefs about 'my' actions.

*"In the end, we are self-perceiving, self-inventing, locked-in mirages that are little miracles of self-reference."* (Hofstadter, 2007) *p.363*.

The notion that internal states encode representations about active states has some nice parallels in neurophysiology in terms of efference copy (Jeannerod and Arbib 2003) and corollary discharges in the brain (Gyr 1972).

**Definition 4**: *strange particles are conservative particles whose active states do not directly influence (i.e., are hidden from) internal states; that is, the flow of internal states in* (10) *does not depend upon active states* $f_\mu(s,a,\mu) = f_\mu(s,\mu)$; *and where* (31) *holds*.

As in the variational free energy lemma, we now construct a generalised free energy whose minima coincide with the internal paths of least action of strange particles:

**Lemma** (*generalised free energy*): the internal paths of a strange particle minimise a free energy functional of Bayesian beliefs about the hidden causes of sensory paths; namely, external and active paths, $q_{\vec{\mu}}(\vec{\eta},\vec{a}) = p(\vec{\eta},\vec{a}\,|\,\vec{s})$:

---

[18] In the case of linear flows in (1), the flow of generalised states in (2) is also linear, which means that the Lagrangian is quadratic, e.g., see Section 6 in (Friston et al. 2022). In this case, the condition under which a variational density can be defined that satisfies (31) is the object of Lemma 2.1 in (Da Costa et al. 2021) using the fact that external and active states are hidden from internal states by the Markov blanket provided by sensory states. Analogously to our reasoning for (non-strange) particles, we expect that in the nonlinear case, for many such systems where sensory states form a Markov blanket, there will be a map affording (31).

$$\vec{\mu} = \arg\min_{\vec{\mu}} \mathcal{L}(\vec{\mu} \mid \vec{s}) \Leftrightarrow \nabla_{\vec{\mu}} \mathcal{L}(\vec{s}, \vec{\mu}) = 0 \Leftrightarrow$$

$$\vec{\mu} = \arg\min_{\vec{\mu}} G(\vec{s}, \vec{\mu}) \Leftrightarrow \nabla_{\vec{\mu}} G(\vec{s}, \vec{\mu}) = 0 \Rightarrow$$

$$\dot{\vec{\mu}} = \mathbf{D}\vec{\mu} - \nabla_{\vec{\alpha}} G(\vec{s}, \vec{\mu})$$

$$\begin{aligned}
G(\vec{s}, \vec{\mu}) &= \underbrace{\mathbb{E}_{q(\vec{\eta},\vec{a})}[\mathcal{L}(\vec{\eta}, \vec{s} \mid \vec{\alpha}) + E(\vec{\alpha}) + \ln q(\vec{\eta}, \vec{a})]}_{\text{Generalised free energy}} \\
&= \underbrace{\mathbb{E}_q[\mathcal{L}(\vec{\eta}, \vec{s}, \vec{a}, \vec{\mu})]}_{\text{Energy constraint}} - \underbrace{\mathbb{E}_q[-\ln q(\vec{\eta}, \vec{a})]}_{\text{Entropy}} \\
&= \underbrace{D_{KL}[q(\vec{\eta}, \vec{a}) \,\|\, p(\vec{\eta}, \vec{a})]}_{\text{Complexity}} + \underbrace{\mathbb{E}_q[\mathcal{L}(\vec{s}, \vec{\mu} \mid \vec{\eta}, \vec{a})]}_{(-) \text{ Accuracy}} \\
&= \underbrace{D_{KL}[q(\vec{\eta}, \vec{a}) \,\|\, p(\vec{\eta}, \vec{a} \mid \vec{s})]}_{\text{Divergence}} + \underbrace{\mathcal{L}(\vec{s}, \vec{\mu})}_{(-) \text{ Log evidence}} \geq \mathcal{L}(\vec{s}, \vec{\mu})
\end{aligned} \tag{32}$$

The generalised free energy expresses a generative model of sensory and internal dynamics in terms of their causes, which include active paths. Choosing internal paths that minimise generalised free energy maximises the evidence afforded to the generative model by sensory paths. This has been referred to as self-evidencing (Hohwy, 2016). Heuristically, a strange particle will look as if it is garnering evidence for its generative model, where its generative model entails the prior belief that it acts like a conservative particle (because it is).

**Proof**: from the definition in **Error! Reference source not found.** the Lagrangian and generalised free energy share the same minima on internal paths of least action, where their gradients vanish:

$$\begin{aligned}
G(\vec{s}, \vec{\mu}) &= \underbrace{D_{KL}[q_{\vec{\mu}}(\vec{\eta}, \vec{a}) \,\|\, p(\vec{\eta}, \vec{a} \mid \vec{s})]}_{=0} + \mathcal{L}(\vec{s}, \vec{\mu}) = \mathcal{L}(\vec{s}, \vec{\mu}) \\
&\Rightarrow \\
\dot{\vec{\mu}}(\tau) &= \mathbf{D}\vec{\mu} - \nabla_{\vec{\mu}} \mathcal{L}(\vec{s}, \vec{\mu}) \\
&= \mathbf{D}\vec{\mu} - \nabla_{\vec{\mu}} F(\vec{s}, \vec{\mu})
\end{aligned} \tag{33}$$

This means we can replace the Lagrangian gradients in **Error! Reference source not found.** with generalised free energy gradients to give (32)

**Remarks**: For strange particles, there is distinction between active paths as *realised variables* and the active paths in (32) as *random variables* that are inferred. This means that the actions of an agent are distinct from beliefs about action, which are based on expected free energy. When the generative model is supplemented with these priors, the ensuing functional corresponds to (upper bound on) the Lagrangian or *surprisal* of sensory outcomes. In short, sensory surprisal is a functional of Bayesian beliefs (i.e., a variational density) about the hidden causes of sensations; namely, external and active dynamics. Internal dynamics minimise this surprisal, while active paths can be described as realising the sensory consequences of inferred action. In effect, the agent authors her sensorium, based upon prior beliefs about the way she acts.

Note that only the internal paths minimise generalised free energy. Nothing has changed from the perspective of the active paths, which can be cast as gradient flows on variational free energy. However, from the perspective of an observer, behaviour will appear to be fulfilling the epistemic and pragmatic imperatives afforded by expected free energy. From a statistical perspective, this corresponds to a move from *control* as inference (Tschantz et al., 2022) to *planning* as inference (Attias, 2003; Botvinick and Toussaint, 2012; Mirza et al., 2016). From a biological perspective, this corresponds to a move from *homeostasis* (Cannon, 1929) to *allostasis* (Ashby, 1947; Corcoran et al., 2020; Ramsay and Woods, 2014; Seth and Friston, 2016; Sterling and Eyer, 1988). Clearly, these distinctions depend upon the time over which path integrals are taken.

We have been deliberately vague about the timescale (or order of generalised motion) over which the expected free energy applies. This vagueness (Machina, 1976) admits a range of temporal horizons for conservative particles: some may have a myopic generative model and implicitly consider path integrals over short periods of time, responding to external fluctuations in a largely reflexive, homeostatic manner (e.g., chemotaxis). Others may have a deep time horizon and exhibit more adaptive, allostatic behaviours (e.g., curiosity). These behaviours are usually associated with deep generative models that feature a separation of temporal scales: for *in silico* examples, please see (Friston et al., 2017d; George and Hawkins, 2009).

## Simulating sentience

This formulation of strange particles can be used to simulate sentient behaviour by solving the following equations of motion yeah, in which internal dynamics minimise generalised free energy and active dynamics minimise the ensuing variational free energy. Combining (21) with (32):

$$\begin{bmatrix} \dot{\vec{\eta}} \\ \dot{\vec{s}} \\ \dot{\vec{a}} \\ \dot{\vec{\mu}} \end{bmatrix} = \begin{bmatrix} \mathbf{f}_{\vec{\eta}}(\vec{\eta}, \vec{s}, \vec{a}) + \vec{\omega}_{\eta} \\ \mathbf{f}_{\vec{s}}(\vec{\eta}, \vec{s}, \vec{a}) + \vec{\omega}_{s} \\ \mathbf{D}\vec{a} - \nabla_{\vec{a}} F(\vec{s}, \vec{a}, \vec{\mu}) \\ \mathbf{D}\vec{\mu} - \nabla_{\vec{\mu}} G(\vec{s}, \vec{\mu}) \end{bmatrix} \quad (34)$$

Figure 5 provides an example in which the expected free energy $E(\vec{\alpha}) = \mathcal{L}(\vec{\alpha})$ was specified directly with a Lagrangian over autonomous paths, in terms of some equations of motion, c.f., a central pattern generator. When specifying generative models like this, the Lagrangian over autonomous paths is usually specified as a prior over *exogenous causes* that intervene in external dynamics. Intuitively—unbeknown to the model—action fulfils the predictions of some prior beliefs about causes that are exogenous to the external dynamics. In other words, even though the particle is causing its sensations, it just 'thinks' it has the right prior beliefs about exogenous dynamics.

As is common in these simulations, exogenous dynamics prescribe itinerant behaviour in the form of a strange attractor. Equation (34) has been used to emulate many kinds of sentient behaviour, ranging from morphogenesis (Friston et al., 2015), through action observation (Friston et al., 2011) to birdsong (Isomura et al., 2019).

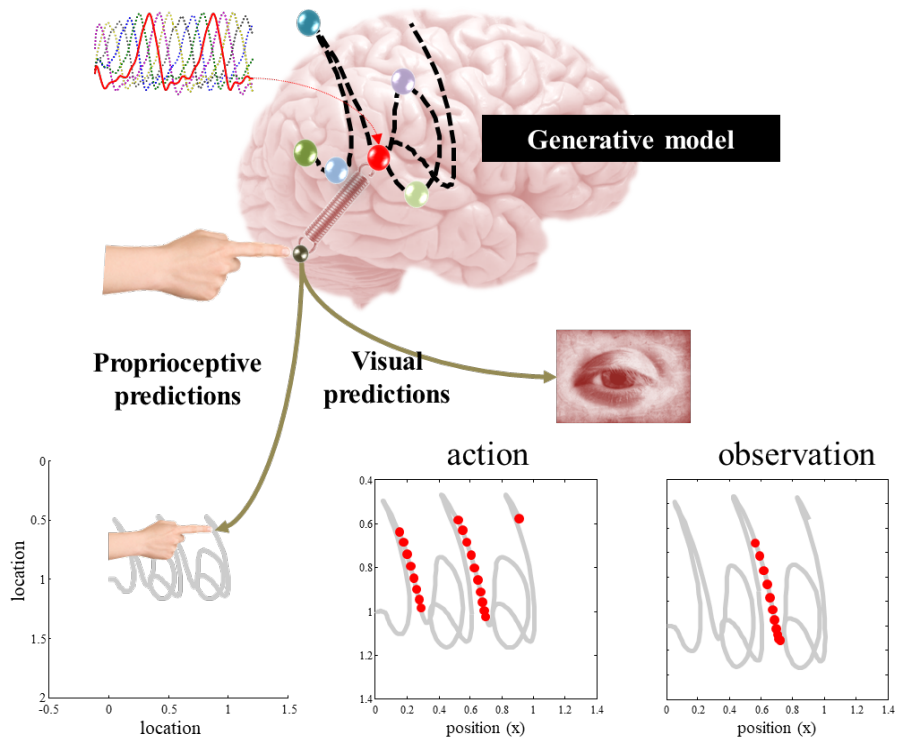

**Figure 5** – *sentient behaviour and action observation*. This figure illustrates simulations of active inference (here, writing), in terms of conditional expectations about hidden states of the world, consequent predictions about sensory input and ensuing action. The dynamics that underwrite this behaviour rest upon prior expectations about exogenous causes that follow Lotka-Volterra dynamics: these are the six (arbitrarily) coloured lines in the upper left inset. In this generative model, each state is associated with a location in Euclidean space that attracts the agent's finger. In effect, the (generalised) internal states then supply predictions of what (generalised) sensory states should register if the agent's beliefs were true. Active states suppress the ensuing prediction error (i.e., maximise accuracy) by fulfilling expected changes in sensed angular velocity, through exerting forces on the agent's joints (not shown). The subsequent movement of the arm is traced out in the lower left panel. This trajectory is plotted in a moving frame of reference, so that it looks like handwriting (e.g., a succession of 'j' and 'a' letters). The lower left panels show the activity of the fourth hidden state under 'action', and 'action-observation'. During action, sensory states register the visual and proprioceptive consequences of movement, while under action observation, only visual sensations are available—as if the agent was watching another agent. The red dots correspond to the times during which this state exceeded an arbitrary threshold. They key thing to note here is that this internal state responds preferentially when, and only when, the motor trajectory produces a down-stroke, but not an up-stroke more—evincing a cardinal feature of neuronal responses, namely, directional selectivity. Furthermore, with a slight delay, this internal state responds during action *and action observation*. From a neurobiological perspective, this is interesting because it speaks to an empirical phenomenon known as mirror neuron activity (Gallese and Goldman, 1998; Kilner et al., 2007; Rizzolatti and Craighero, 2004). Please see (Friston et al., 2011) for further details.

The handwriting example in Figure 5 illustrates a simple sort of active inference specified in terms of a Lagrangian $\mathcal{L}(\vec{\alpha})$ over autonomous (i.e., exogenous) causes. To simulate the curious behaviour of strange particles like ourselves, one can evaluate expected free energy explicitly, $E(\vec{\alpha}) = \mathcal{L}(\vec{\alpha})$.

Effectively, this enables a specification of autonomous dynamics in terms of Lagrangians (i.e., equations of motion) that engender information and preference seeking. Practically speaking, knowing the functional form of expected free energy allows one to simulate or reproduce sentient behaviour that is specified in terms of preferred outcomes. The requisite free energy functionals are:

$$E(\vec{\alpha}) = \underbrace{\mathbb{E}_{p(\vec{s}|\vec{\eta},\vec{\alpha})q(\vec{\eta}|\vec{\alpha})}[\mathcal{L}(\vec{\eta},\vec{s}) - \mathcal{L}(\vec{\eta}\,|\,\vec{\alpha})]}_{\text{Expected free energy}} = \mathcal{L}(\vec{\alpha})$$

$$F(\vec{s},\vec{\alpha}) = \underbrace{\mathbb{E}_{q(\vec{\eta})}[\mathcal{L}(\vec{\eta},\vec{s},\vec{\alpha}) + \ln q(\vec{\eta})]}_{\text{Variational free energy}} \geq \mathcal{L}(\vec{s},\vec{\alpha})$$

$$G(\vec{s},\vec{\mu}) = \underbrace{\mathbb{E}_{q(\vec{\eta},\vec{a})}[\mathcal{L}(\vec{\eta},\vec{s}\,|\,\vec{\alpha}) + E(\vec{\alpha}) + \ln q(\vec{\eta},\vec{a})]}_{\text{Generalised free energy}} \geq \mathcal{L}(\vec{s},\vec{\mu}) \quad (35)$$

$$F(\vec{s},\vec{\mathbf{\alpha}}) = \mathcal{L}(\vec{s},\vec{\mathbf{\alpha}}) : \vec{\mathbf{a}} = \arg\min_{\vec{a}} F(\vec{s},\vec{a},\vec{\mathbf{\mu}})$$

$$G(\vec{s},\vec{\mathbf{\mu}}) = \mathcal{L}(\vec{s},\vec{\mathbf{\mu}}) : \vec{\mathbf{\mu}} = \arg\min_{\vec{\mu}} G(\vec{s},\vec{\mu})$$

The expected free energy above is an *expectation* under a predictive density over hidden causes and sensory consequences, based on beliefs about external states, supplied by the variational density. Intuitively, based upon beliefs about the current state of affairs, the expected free energy furnishes the most likely 'direction of travel' or path into the future. This construction specifies the most likely behaviour in terms of expected outcomes, where the expected free energy brings a belief-dependent or epistemic aspect to behaviour; namely, *curiosity* (Friston et al., 2017b; Schmidhuber, 2006; Still and Precup, 2012).

This lends a complementary meaning to the *expected* free energy; in the sense it is expected in the future (and past). This implicit prospection (and postdiction) means that sensory paths into the future (and past) are random variables. In short, strange particles or agents (look as if they) think they are conservative particles, and act accordingly. The example in Figure 6 suppresses prior preferences to reveal pure information-seeking or epistemic behaviour—a succession of visual palpations—in the setting of visual search and scene construction. This example specified a sequence of active trajectories to fixed points, specified as exogenous causes. These visual fixation points minimised expected free energy; i.e., maximised expected information gain.

In summary, strange particles look as if they have agency, in the sense their actions realise the predictions of Bayesian beliefs. This leads to a slightly counterintuitive interpretation of the ensuing behaviour, in which active states are hidden from internal representations of action. Intuitively, this means beliefs about how the world should play out are realised, reflexively, by active states. From the perspective of someone observing an agent, say a fish, it will look as if the fish searches out particles of food. However, from the point of view of the fish, it believes that it is propelled through water in a fortuitous and benevolent way that delivers food particles to its mouth. In other words, the fish is unaware it is the agent of its actions, it just believes this is how the world works—beliefs that are realised through action: cf., ideomotor and perceptual control theory (Mansell, 2011; Pfister et al., 2014; Seth, 2015; Wiese, 2017). Note that this is not inconsistent with the fish learning the association between specific internal (e.g., neural) stimuli and the sensory consequences of action, thus becoming aware, that it can control its own actions—or at least their consequences.

This speaks to further particular kinds, with deeper generative models, that (look as if they) recognise action is underwritten by agency. Or indeed, they are the authors of their actions. These kinds of agents may be distinguished by generative models of state-dependent random fluctuations that introduce further sort of hierarchical depth; namely, beliefs about beliefs in the form of beliefs about uncertainty or precision[19] (Clark, 2013a; Limanowski, 2017; Limanowski and Blankenburg, 2013; Seth, 2013).

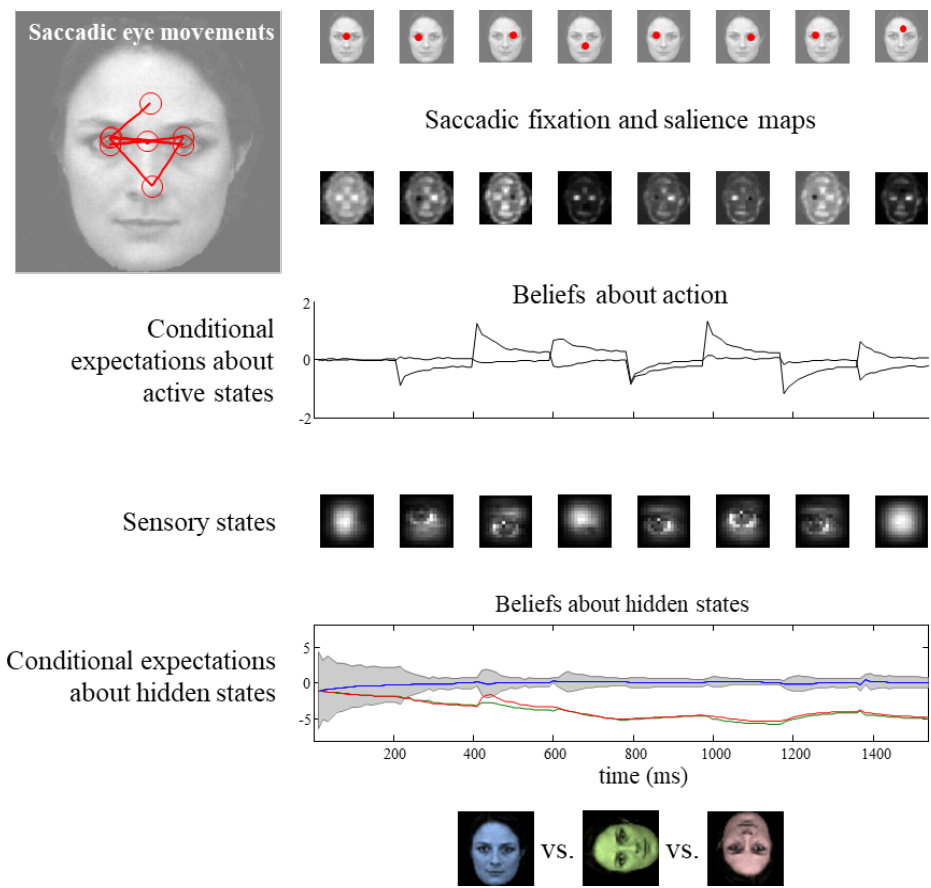

**Figure 6** – *Epistemic foraging*. This figure shows the results of a simulation in which a face was presented to an agent, whose responses were simulated by selecting actions that minimised expected free energy following an eye movement. The agent had three internal beliefs or hypotheses about the stimuli she might sample (an upright face, an inverted face and a rotated face). The agent was presented with an upright face and her posterior expectations were evaluated over 16 (12 ms) time bins, until the next saccade was emitted. This was repeated for eight saccades.

---

[19] That is, the inverse covariances in (19).

The ensuing fixation points are shown as red dots in the upper row. The corresponding sequence of eye movements is shown in the inset on the upper left, where the red circles correspond roughly to the proportion of the visual image sampled. These saccades are driven by predictive beliefs about the next fixation point based upon salience maps in the second row. These salience maps are the (negative) expected free energy as a function of action; namely, where to look next. Note that these maps change with successive saccades as posterior beliefs about hidden causes become more confident. Note also that salience is depleted in locations that were foveated in the previous saccade, because these locations no longer have epistemic affordance (i.e., the ability to reduce uncertainty or expected information gain). Empirically, this is known as inhibition of return. Oculomotor responses are shown in the third row in terms of the two hidden oculomotor states corresponding to vertical and horizontal eye movements. The associated portions of the image sampled (at the end of each saccade) are shown in the fourth row. The final two rows show the accompanying posterior beliefs, in terms of posterior expectations and 90% Bayesian credible intervals. This illustrates the nature of active inference when sampling data to disambiguate among hypotheses or percepts that best explain sensory evidence. Please see (Friston et al., 2012) for further details.

## Epilogue: on the nature of strangeness

"*Now, here, you see, it takes all the running you can do, to keep in the same place*", the Red Queen to Alice in Lewis Carroll's Through the Looking-Glass.

Clearly, we have taken some poetic licence in associating the recursion implicit in the dynamics of strange particles with a strange loop (Hofstadter, 2007). However, there exists a complementary perspective on conservative dynamics that brings us back to strange loops. This is nicely exemplified by Red Queen dynamics (Ao, 2005; Zhang et al., 2012), which can be read as: moving forwards "one winds up, to one's shock, exactly where one had started out". Red Queen dynamics are one of many formulations of conservative, divergence-free, or solenoidal flow that underwrites nonequilibrium steady-states and the implicit breaking of detailed balance that characterises living systems (Ao, 2008; Haken, 1983; Nicolis and Prigogine, 1977).

To see why this kind of dynamics is characteristic of conservative particles, we turn to a complementary decomposition of the flow afforded by a Helmholtz decomposition (Ao, 2004; Barp et al., 2021; Da Costa and Pavliotis, 2022; Eyink et al., 1996; Graham, 1977a; Ma et al., 2015; Shi et al., 2012). This decomposes the flow of *states* into a *conservative* solenoidal component and a *dissipative* gradient flow that depends upon the amplitude of random fluctuations

$$f(x) = \underbrace{Q(x)\nabla \Im(x)}_{\text{Conservative}} - \underbrace{\Gamma(x)\nabla \Im(x)}_{\text{Dissipative}} - \underbrace{\nabla \cdot Q(x) + \nabla \cdot \Gamma(x)}_{\text{correction terms}}. \qquad (36)$$

Here, $\Im(x) = -\ln p(x)$ is the self-information or surprisal under the (nonequilibrium steady-state) solution to the density dynamics in (1). The sensorimotor loop[20]—implicit in our definition of particle—ensures the presence of solenoidal flow, which breaks time-reversal symmetry (Jiang et al., 2004).

When random fluctuations are small, the solenoidal contribution predominates, leading to chaotic itinerancy turbulence, solenoidal mixing and strange attractors (Friston et al., 2021; Namikawa, 2005; Parr et al., 2020; Pavlos et al., 2012; Takens, 1980; Tsuda and Fujii, 2004). Solenoidal flow is conservative because it circulates on the level sets of the Lagrangian. This licenses the name *conservative* particles, as conservative flow governs the motion of particular states. In the limit of vanishingly small fluctuations, we have an attracting set of states to which the paths are confined. In other words, trajectories will always bring systemic states back to the neighbourhood of a previously occupied state; namely, ending up "where one had started out". It is interesting to note that most of the biomimetic simulations referenced above are based on a Lagrangian (i.e., generative model) whose equations of motion have a strange attractor—an attracting set with a fractional dimension (Ma et al., 2014); e.g., (Friston and Frith, 2015). These attractors feature recurrent loops through Poincaré sections (Bramburger and Kutz, 2020) that may qualify as a strange loop, of a sort.

---

[20] By sensorimotor loop we mean that sensory states influence internal states that in turn influence active states, and not *vice versa*.

## Discussion

"*Was it utterly absurd to seek behind the ordering structures of this world a 'consciousness' whose 'intentions' were these very structures?*" (Heisenberg, 1971)

There are many issues that attend the free energy principle. We take this opportunity to briefly consider four.

The first is a link to early formulations of self-organisation in cybernetics; namely, Ashby's Law of Requisite Variety (Ashby, 1956) that underpins the Good Regulator Theorem (Conant and Ashby, 1970), later articulated as the Internal Model Principle of control theory (Francis and Wonham, 1976). The Law of Requisite Variety pertains to the degrees of freedom of an agent's internal model. Put simply, an organism (i.e., active particle) must have a repertoire of states that is at least equal to the number of fluctuations in the environment (i.e., external states). As often quoted: "only variety can absorb [destroy] variety". In the present setting, this speaks to the number of generalised internal states, in relation to the number of generalised external states that matter. External states that matter are slow (dynamically unstable) states that can be distinguished from fast (dynamically stable) fluctuations (Carr, 1981; Frank, 2004; Haken, 1983; Koide, 2017). The Law of Requisite Variety emerges under the above formalism from the requirement that the number of generalised internal states must be greater than the number of generalised external states (that matter). This follows because internal states play the role of sufficient statistics of a variational density over generalised external states, with one or more statistic for each external state. This instance of the Law of Requisite Variety speaks to the vague nature of sentient behaviour described by the free energy principle.

Sentience is read here as an attribute of behaviour that rests on inference; namely, movement on a statistical manifold, on which internal states evolve (Parr et al. 2020; Friston et al. 2022). Movement on this (statistical) manifold corresponds to belief updating and thereby an elementary kind of sentience or sense making with a well-defined information geometry (Caticha, 2015; Ikeda et al., 2004; Kim, 2018). Having said this, sentient behaviour can be so elemental as to be trivial. For example, a thermostat with one degree of freedom to act upon— and represent—its external milieu can only exhibit a weak sort of sentient behaviour from the perspective of a particle (like you and me) with a more expressive model of the thermostat and its external states. The FEP makes a clear commitment to describing sentience as embodied cognition, where embodiment is supplied by the interaction—and implicit boundary—between a 'thing' and everything else, formulated as dynamical coupling via a Markov blanket.

This touches on our second issue; namely, any empirical and philosophical claims about the free energy principle, in relation to sentient behaviour: in brief, the free energy principle is a method for describing the behaviour of certain kinds of particles that may have strong or weak sentience, depending upon the way in which one particle (e.g., you or me) makes sense of another (e.g., a thermostat). These attributes of strong or weak sentience emerge as a consequence of the interaction between the various states of a particle and the states external to it. It may be possible to experimentally demonstrate where any given particle (e.g., molecules, mice, and men) belongs in this sentient hierarchy by simply examining the nature of their particular dynamics. For example, this treatment of the FEP rests upon the smooth nature of fluctuations in (1), cf. (Friston et al. 2022), and the nature of the coupling in (10), both of which may be empirically tested via Bayesian selection of stochastic dynamical models given time-series data from a particle and its surrounding external states. On a more practical level, there are procedures for identifying nested Markov blankets (across scales) from empirical time-series. Please see (Friston et al. 2020) for a worked example using brain imaging data. Whether these procedures find purchase in decomposing distributed systems across scales, in empirical studies, remains to be seen.

When the world external to a particle is constituted of other particles, we enter the realm of collective behaviour and interacting particles. From the perspective of active matter, the key perspective offered by the FEP is the Markov blanket that segregates the internal degrees of freedom of a particle from other particles. It implies that agents implicitly have beliefs about other agents—that is, generative models that entail a sense of others, and perhaps a sense of self. Interactions between agents occur through the blanket in a process of active inference—not solely by mechanical forces—introducing a sentient aspect to interactions. The distinction between various particular kinds may be particularly apt to account for different kinds of active matter—from least to most sentient—as is encountered in varying degrees throughout the life sciences. It would be interesting to analyse the extent to which ensembles of increasingly sentient particles feature richer interactions (e.g., communication) among agents, in relation to the variously studied collective phenomena (e.g., pattern formation, flocking, and phase transitions). In machine learning, where interacting particles are often used for the purpose of achieving a goal faster than a single particle could (Borovykh et al., 2021), introducing more sentient (i.e., beyond inert) particles may also lead to richer interactions and more 'intelligent' behaviour: from biological intelligence to distributed cognition (Levin, 2019).

Third, from a statistical perspective, there are interesting parallels between applications of the free energy principle and the complete class theorem (Brown, 1981; Wald, 1947). The complete class theorem asserts that for any pair of behaviours and preferences, there are some priors that render the behaviour Bayes optimal. This translates, in the setting of the free energy principle, into the assertion that there exists some internal dynamics, which encode prior beliefs about external dynamics that render autonomous dynamics Bayes optimal. Whether this description of autonomous behaviour is an apt description of internal dynamics *per se* is a question that cannot be answered. This is because internal states are inaccessible, by construction—they are private to the particle in question. This leaves one in the game of inferring internal dynamics on the basis of sensory and active trajectories, which is a fair description of most of the life sciences; especially, the neurosciences: e.g., (Zeki and Shipp, 1988). For example, much of modern neuroscience attempts to peer through the brain's Markov blanket, using neuroimaging and electrophysiology. Or, breaching the Markov blanket to observe internal states, through the use of post-mortem studies and invasive procedures. Interestingly, much of computational neuroscience can be framed as finding the generative model that best explains a subject's choices and responses (Parr et al., 2018).

Finally, it is worth noting many things have not been addressed in this brief account of the free energy principle. For example, we have ignored states that endow dynamics with a functional form; namely, the parameters of the equations of motion that define a Lagrangian. These parameters necessarily introduce a separation of temporal scales, in the sense that the parameters of the flow operators change slowly in relation to states. Anecdotally, this introduces a distinction between *inference* and *learning* over fast and slow timescales, respectively. At this level of analysis there are many outstanding issues (Fields et al., 2021b). For example, how does inference underwrite learning? Can autopoiesis (Maturana and Varela, 1980) be understood in terms of blanket-building at a slow timescale? Is there a natural progression over scales to more conservative (and strange) mechanics (Jeffery et al., 2019)? How do particles of particles behave (Heins et al., 2022; Kuchling et al., 2020; Levin, 2019; Palacios et al., 2020) and so on…

## Additional Information

**Funding Statement**

KF is supported by funding for the Wellcome Centre for Human Neuroimaging (Ref: 205103/Z/16/Z) and a Canada-UK Artificial Intelligence Initiative (Ref: ES/T01279X/1). L.D. is supported by the Fonds National de la Recherche, Luxembourg (Project code: 13568875). This publication is based on work partially supported by the EPSRC Centre for Doctoral Training in Mathematics of Random Systems: Analysis, Modelling and Simulation (EP/S023925/1). CH is supported by the U.S. Office of Naval Research (N00014-19-1-2556). This research was funded in part by the Wellcome Trust [205103/Z/16/Z]. For the purpose of Open Access, the author has applied a CC BY public copyright license to any Author Accepted Manuscript version arising from this submission. The work of GAP was partially funded by the EPSRC, grant number EP/P031587/1, and by JPMorgan Chase & Co through a Faculty Research Award 2019 and 2021.

**Acknowledgements**

We would like to thank Samuel Tenka, Mel Andrews, and the co-organisers of the International Physics Reading Group with Maxwell Ramstead, who generated many of the issues and questions addressed in this paper. We thank our anonymous reviewers for their helpful comments which improved the manuscript.

**Competing Interests**

The authors have no competing interests.

**Authors' Contributions**

All authors made substantial contributions to conception and design, and writing of the article, and approved publication of the final version.